\documentclass[usenatbib]{mn2e}
\usepackage{graphicx}

\def\apj{\rm ApJ}
\def\apjl{\rm ApJL}
\def\apjs{\rm ApJS}
\def\aj{\rm AJ}
\def\mnras{\rm MNRAS}
\def\nat{\rm Nature}
\def\pasj{\rm PASJ}
\def\pasp{\rm PASP}
\def\pasa{\rm PASA}
\def\aap{\rm AAP}
\def\araa{\rm ARA\&A}

\def\apss{\rm Ap\&SS}

\def\gax{\mathrel{\raise.3ex\hbox{$>$}\mkern-14mu\lower0.6ex\hbox{$\sim$}}}
\def\lax{\mathrel{\raise.3ex\hbox{$<$}\mkern-14mu\lower0.6ex\hbox{$\sim$}}}
\def\gtorder{\mathrel{\raise.3ex\hbox{$>$}\mkern-14mu
             \lower0.6ex\hbox{$\sim$}}}
\def\ltorder{\mathrel{\raise.3ex\hbox{$<$}\mkern-14mu
             \lower0.6ex\hbox{$\sim$}}}

\begin{document}

\title [Stellar Binaries That Survive Supernovae]
   {Stellar Binaries That Survive Supernovae}

\author[C.~S. Kochanek et al.]{ 
    C.~S. Kochanek$^{1,2}$, 
    K. Auchettl$^{3}$ and 
    C. Belczynski$^{4}$ 
    \\
  $^{1}$ Department of Astronomy, The Ohio State University, 140 West 18th Avenue, Columbus OH 43210 \\
  $^{2}$ Center for Cosmology and AstroParticle Physics, The Ohio State University,
    191 W. Woodruff Avenue, Columbus OH 43210 \\
  $^{3}$ Dark Cosmology Centre, Niels Bohr Institute, University of Copenhagen, 
      Juliane Maries Vej 30, 2100, Copenhagen, Denmark \\
  $^{4}$ Nicolaus Copernicus Astronomical Center, Polish Academy of Sciences,
           ul. Bartycka 18, 00-716 Warsaw, Poland
   }

\maketitle

\begin{abstract}
The number of binaries containing black holes (BH) or neutron stars (NS) depends 
critically on the fraction of binaries that survive supernova (SN) explosions.
We searched for surviving star plus remnant binaries in a sample of 49 supernova remnants (SNR)
containing 23 previously identified compact remnants and three high mass X-ray
binaries (HMXB), finding no new interacting or non-interacting binaries.  The
upper limits on any main sequence stellar companion are typically $\ltorder 0.2M_\odot$
and are at worst $\ltorder 3M_\odot$. This implies that 
$f < 0.1$ of core collapse SNRs contain a non-interacting binary, and $f=0.083$ 
($0.032 < f < 0.17$) contain an interacting binary at 90\% confidence. We also
find that the transverse velocities of HMXBs are low, with a median of only
$12$~km/s for field HMXBs, so surviving binaries will generally be found very
close to the explosion center.  We compare the results to a 
``standard'' {\tt StarTrack} 
binary population synthesis (BPS) model, finding reasonable agreement with the
observations.  In particular, the BPS models predict that 5\% of SNe should 
leave a star plus remnant binary.
\end{abstract}

\begin{keywords}
stars: massive -- supernovae: general -- supernovae
\end{keywords}

\section{Introduction}

Most massive stars are in binaries (see, e.g., \citealt{Sana2012},
\citealt{Duchene2013}, \citealt{Kobulnicky2014} and \citealt{Moe2016}).
For example, \cite{Moe2016} find that only $16\pm8$\% of stars that
dominate the core-collapse supernova (ccSN) rate are single.  Binary evolution,
through mass transfer, mass loss, and merging, modifies the binary
population as the stars evolve (e.g., \citealt{Eldridge2008},
\citealt{Sana2012}, \citealt{Renzo2018}, \citealt{Zapartas2017}). 
Then, when the primary 
explodes, the binary can become unbound either due to mass loss
(\citealt{Blaauw1961}) or ``kicks'' due to the explosion
(e.g., \citealt{Gunn1970}, \citealt{Iben1996}, \citealt{Cordes1998}, 
\citealt{Faucher2006}).  For the binaries which survive, 
further evolution can lead to high and low mass X-ray binaries
(HMXBs and LMXBs, see, e.g., the reviews by \citealt{Remillard2006},
\citealt{Reig2011}, or \citealt{Walter2015}), and the explosion (or collapse) of the
secondary can lead to the formation of neutron star (NS) and black
hole (BH) binaries that can be gravitational wave sources (e.g., \citealt{Abbott2016}, \citealt{Abbott2017}).
This means that the fraction of binaries which survive ccSNe
and their properties are crucial ingredients for any process
involving NS or BH binaries.

A simple toy model assuming passively evolving binaries (i.e.,
no interactions) can help frame the basic statistics (\citealt{Kochanek2009}).  
We can view the effective binary fraction as $F = F_0 (1-f_m) \simeq 55\%$, where
$F_0 \simeq 84\%$ (\citealt{Moe2016}) is the true initial binary fraction 
and $f_m$ is the fraction of binaries which merge before a ccSN can occur.
In our binary population synthesis (BPS) model (\S\ref{sec:theory}), we 
find $f_m=34\%$, while \cite{Renzo2018} find $f_m \simeq 22_{-9}^{+26}\%$.
The fractions of ccSNe which occur in stellar binaries, as a single
stars (including merger remnants), and as explosions of a 
secondary (which may be in a binary with a compact remnant) are
\begin{equation}
    f_b = { F \over 1 + F f_q },
    \quad
   f_1 = { 1 - F \over 1 + F f_q } 
   \quad\hbox{and}\quad
   f_s = { F f_q \over 1 + F f_q },
\end{equation}
respectively, where    
$f_q = \int_0^1 q^{x-1} P(q) dq$,
$x \simeq 2.35$ is the slope of a \cite{Salpeter1955} initial mass function (IMF),
$0 \leq q = M_2/M_1 \leq 1$ is the mass ratio and
$P(q)$ with $\int_0^1 dq P(q) \equiv 1$ is the distribution of mass
ratios. For constant $P(q)$ (see the discussions in \citealt{Kobulnicky2014},
\citealt{Moe2016}) from $q=0$ to $q=1$ for simplicity, the integral is 
$f_q=0.426$.  Given these assumptions, fraction
$f_b \simeq 41\%$ to $57\%$ of ccSNe occur in stellar binaries for 
effective binary fractions of $F=50\%$ to $75\%$.

At the time of explosion, most stellar companions are relatively 
massive.  If stars more massive than $M_{SN}$ explode,
then the fraction of companions above mass $M_s$ in our simple model
is $ 1 - (x-1)M_s/x M_{SN} \simeq 1- 0.43 M_s/M_{SN}$ for $M_s < M_{SN}$.
If $M_{SN}=8 M_\odot$, 43\% of stellar secondaries are massive enough to 
eventually explode, and 93\%, 64\% and 50\% are more massive than
$M_s > 1$, $3$ and $5M_\odot$, respectively.  Unlike searches for
surviving single degenerate companions to Type~Ia SNe (e.g., \citealt{Schweizer1980},
\citealt{Ruiz2004}, \citealt{Ihara2007}, \citealt{Gonzalez2012},
\citealt{Schaefer2012}), searches for stellar companions to 
ccSNe can focus on relatively high mass, luminous stars with only modest 
statistical penalties.  

If we ignore kicks, the binary becomes unbound if the final system mass
is less than one-half of the initial mass (\citealt{Blaauw1961}). 
If the primary, secondary and neutron star (NS) remnant have masses
of $M_p$, $M_s$ and $M_{ns} \simeq 1.4M_\odot$, respectively, then the secondary 
must have $M_s > M_p - 2M_{ns}$ for the binary to survive. 
If the minimum mass for a ccSN is $M_{SN} \simeq 8 M_\odot$, then
only binaries with secondaries more massive than $M_s  \gtorder 5 M_\odot$
can survive the formation of a NS. This means that in 
our non-interacting binary model, the fraction of binaries that survive the 
explosion of the primary is $2(x-1) M_{NS}/x M_{SN} \simeq 20\%$.  
For effective binary fractions of $F=50\%$ to 75\%, 
$\sim 10\%$ of core collapse SNRs should contain a star-NS binary and 30-45\% 
of SNRs should contain a disrupted binary.  The stars in disrupted 
binaries should, on average, be less massive than the ones
still in binaries.  These back of the envelope 
estimates come surprisingly close to the results of full population 
synthesis models. In \S2 we find that 25\% of stellar binaries survive
the explosion in our BPS models, while  \cite{Renzo2018} find that 
$14_{-10}^{+22}\%$ of binaries survive, but do not distinguish
between companion types.

Searches for binary companions to ccSNe are largely restricted to
the Galaxy and potentially the Magellanic Clouds.  For ccSN in 
nearby ($1$-$30$~Mpc) galaxies, there are two problems.  First,  only the most 
luminous secondaries can be detected (see \citealt{Kochanek2009}).  
The typical main sequence companion is too 
faint to be observed.  Second, if ccSNe form dust in their ejecta
(as seen, e.g., for SN~1987a, \citealt{Matsuura2015}), 
then the secondary will also be heavily dust obscured for decades 
(\citealt{Kochanek2017}).  There are several possible detections
of surviving secondaries
(e.g., SN~1993J, \citealt{Maund2004}, \citealt{Fox2014},
SN~2011dh, \citealt{Folatelli2014} but see \citealt{Maund2015},
and iPTF13bvn, \citealt{Bersten2014}), but it is unlikely that 
distant ccSN can be used to carry out any census of binary 
companions.  Moreover, it will be virtually impossible to 
determine if the binary is still bound.

Most observational efforts to explore the relationship between
ccSNe and binaries have focused on understanding runaway B stars 
(e.g., \citealt{Blaauw1961}, \citealt{Gies1986}, \citealt{Hoogerwerf2001}, 
\citealt{Tetzlaff2011}, \citealt{Renzo2018}) and the 
contribution of binary disruption to the velocities of neutron 
stars (e.g., \citealt{Gunn1970}, 
\citealt{Iben1996}, \citealt{Cordes1998}, \citealt{Faucher2006}). 
Identifying stellar companions has focused on Type~Ia SNe
and the single versus double degenerate problem
(e.g., \citealt{Schweizer1980}
\citealt{Ruiz2004}, \citealt{Ihara2007}, \citealt{Gonzalez2012},
\citealt{Schaefer2012}), with much less attention to finding
stellar companions to ccSNe.

\cite{vandenBergh1980} seems
to have made the first search of supernova remnants (SNRs) for runaway
stars by looking for a statistical excess of O stars close to the 
centers of 17 Galactic SNRs and finding none.  \cite{Guseinov2005} examined 48
Galactic SNRs for O or B stars using simple color, magnitude and 
proper motion selection cuts to produce a list of candidates, but
did not investigate them in detail.  \cite{Dincel2015} identify 
and characterize a candidate unbound binary star in the 
$\sim 3 \times 10^4$~year old SNR G180.0$-$01.7 (S147). This SNR
also contains PSR~J0538+2817, and \cite{Dincel2015} argue that 
the pair were likely a binary before the SN.  
\cite{Kochanek2018} found that the Crab and Cas~A were not binaries
at the time of their explosions, finding no possible former secondaries
down to mass ratios $q \ltorder 0.1$.  The result for Cas~A
was later confirmed by \cite{Kerzendorf2018}. \cite{Boubert2017}
identify 4 candidates, associated with the SNRs G074.0$-$08.5
(Cygnus Loop), G089.0$+$04.7 (HB~21), G180.0$-$01.7 (S147, this
is the same candidate as \citealt{Dincel2015}), and G205.5$+$00.5
(Monoceros Loop).  

Finding stars associated with binaries disrupted by the ccSNe is
challenging because the star is no longer co-located with the
compact object.  Some combination of parallaxes and proper
motions must be used to identify stars that are consistent
with an estimate of the explosion center both spatially and 
temporally.  Unfortunately, accurate stellar parallaxes are 
not a panacea because comparably accurate distances for the compact 
remnants and SNRs are generally lacking.  The interpretation of
proper motions is complicated by the difficulties in determining
explosion centers, as discussed in \cite{Holland2017}.

A much simpler problem is to search for
surviving binaries because the search position is exactly known
from the identification of the compact remnant.  Somewhat to our
surprise, this exercise seems never to have been carried out and
we attempt the first such survey here. This does not mean that 
we will have results free of ambiguities. But we will discuss
these issues as we proceed, and lay out a program to address them
in our conclusions.  We first consider our theoretical expectations
in more detail using BPS models in
\S\ref{sec:theory}.  
Next we select a sample of SNRs from a combination of 
\cite{Ferrand2012} and \cite{Green2014}, and search for
any associated compact objects at radio, X-ray, or $\gamma$-ray
wavelengths.  For the SNRs where compact objects have been
identified, we estimate the stellar masses for the ones which
have a stellar companion and set mass limits for those which
apparently do not. 
This process and its potential selection
effects are discussed in \S\ref{sec:sample}. The properties
of the SNRs with compact objects are discussed in 
Appendix~\ref{sec:accepted}, SNRs without clearly identified
compact objects are 
discussed in Appendix~\ref{sec:nocompact}, and rejected
SNRs are discussed in Appendix~\ref{sec:rejected}.
We also examine the transverse velocities of HMXBs
from the catalog of \cite{Liu2006}
for comparison to the velocities of the binaries and
non-binaries in the SNRs.  We discuss the consequences
for binaries and ccSN in \S\ref{sec:results}, and outline
future possibilities in \S\ref{sec:conclude}

\begin{figure}
\centering
\includegraphics[width=0.50\textwidth]{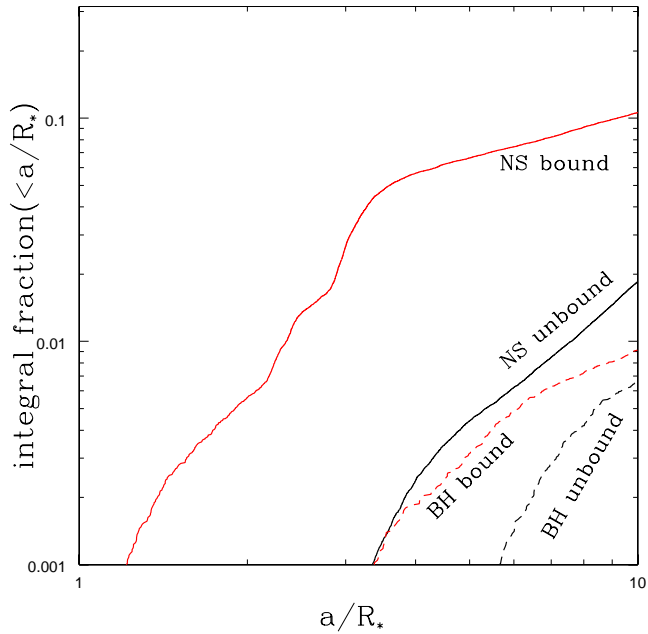}
\caption{ 
  The integral distributions of the ratio $a/R_*$ between
  the pre-SN semimajor axis $a$ and the companion stellar 
  radius $R_*$ divided by whether the core collapse forms
  and NS or a BH and whether the binary is bound or unbound
  afterwards.
  }
\label{fig:edist}
\end{figure}

\begin{figure}
\centering
\includegraphics[width=0.50\textwidth]{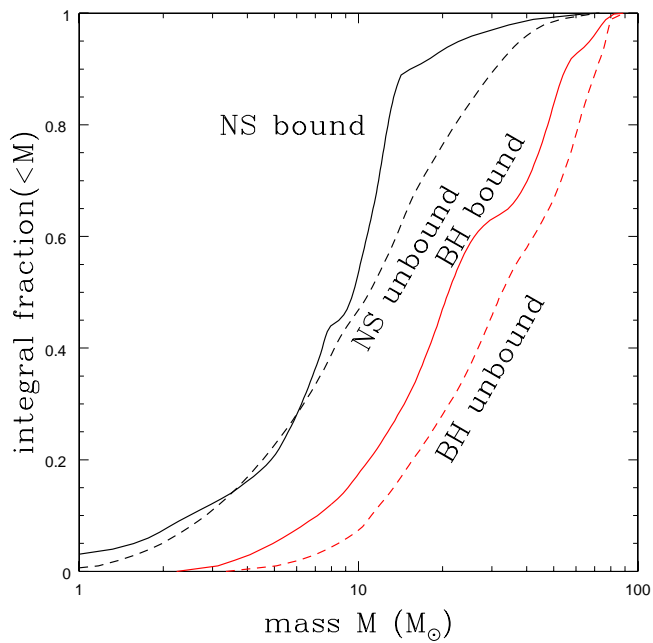}
\caption{ 
  The integral distributions of stellar masses $M$ for
  stellar companions (i.e., excluding WD, NS, BH) at the
  time of core-collapse, separated by the compact object
  formed (NS in black, BH in red) and whether the
  binary remains bound (solid) or unbound (dashed).
  }
\label{fig:mdist}
\end{figure}

\begin{figure}
\centering
\includegraphics[width=0.50\textwidth]{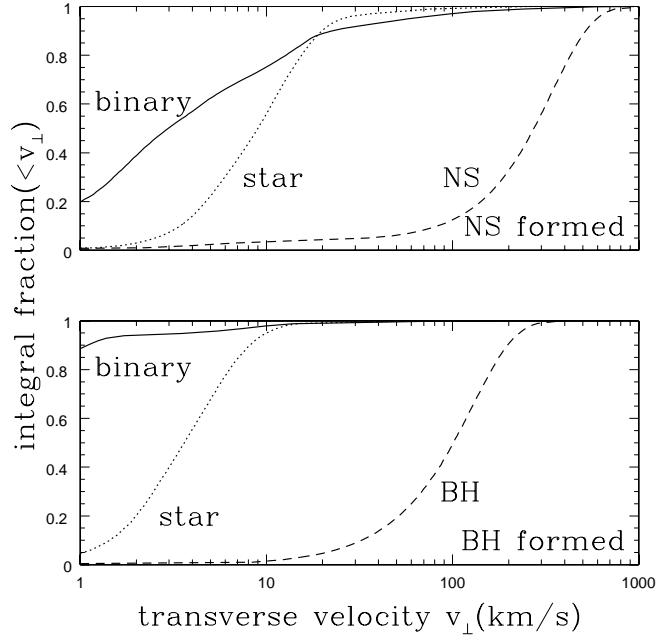}
\caption{ 
  Integral distributions of the transverse velocities $v_\perp$ 
  after core collapse, excluding cases where the companion is
  a compact object, separated by whether a NS (top) or BH (bottom) is formed.
  These velocities are in the rest frame of the center of mass of
  the pre-SN binary.
  The three curves are for surviving binaries (solid, ``binary''), 
  unbound stellar companions (dotted, ``star''), and 
  unbound compact objects (dashed, ``NS'' or ``BH'').
  }
\label{fig:vdist}
\end{figure}

\begin{figure}
\centering
\includegraphics[width=0.50\textwidth]{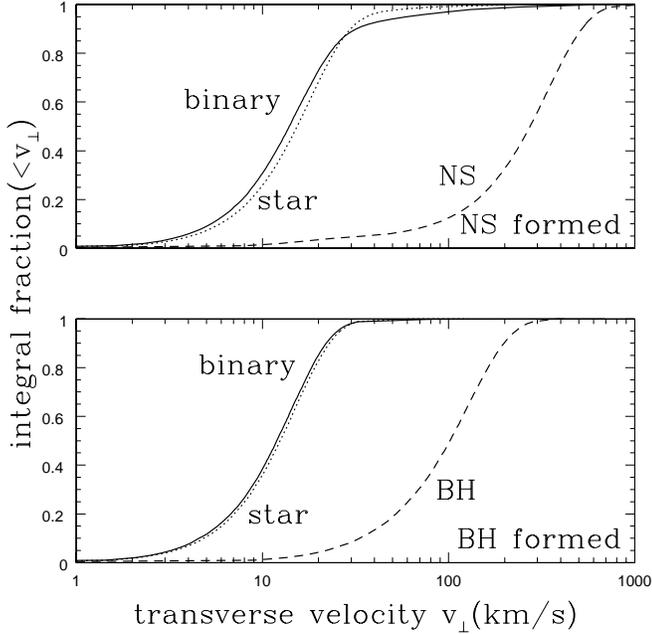}
\caption{ 
  Integral distributions of the transverse velocities $v_\perp$
  measured relative to the mean motion of nearby stars assuming
  a 1D velocity dispersion of $10$~km/s.  This
  excludes cases where the companion is 
  a compact object and is separated by whether a NS (top) or BH (bottom) is formed.
  The three curves are for surviving binaries (solid, ``binary''),
  unbound stellar companions (dotted, ``star''), and
  unbound compact objects (dashed, ``NS'' or ``BH'').
  }
\label{fig:vdist2}
\end{figure}

\begin{figure}
\centering
\includegraphics[width=0.45\textwidth]{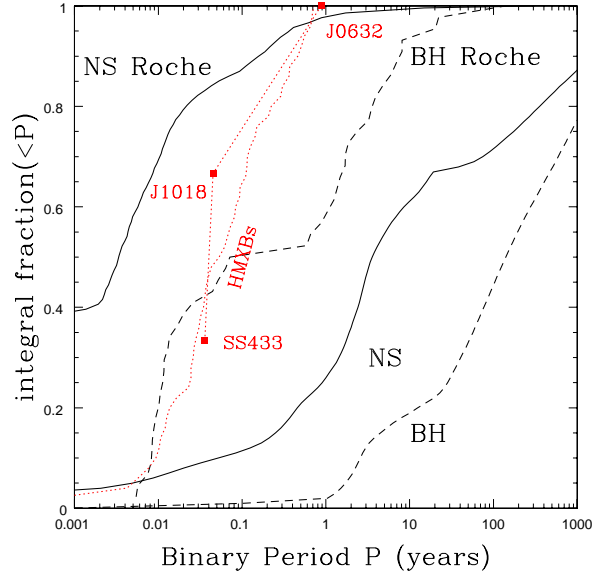}
\caption{ 
  Integral distributions of NS (black solid) and BH (black dashed) stellar
  binary periods after core collapse.  The curves labeled ``NS/BH Roche'' 
  are the period distributions of the binaries for which the orbital
  pericenter lies inside the star's Roche lobe after eliminating actual
  collisions.  The integral period distribution of the three HMXBs associated
  with our sample of SNRs and of the \protect\cite{Liu2006} catalog of 
  field HMXBs
  are also shown.  SS~433 is a Roche lobe accretion system, while
  J0632 and J1018 are wind accretion/interaction systems.  The 
  distribution of the overall HMXB population in period will be
  affected by the post-SN evolution of the star and binary.
  }
\label{fig:bperiod}
\end{figure}

\section{Theoretical Expectations}
\label{sec:theory}

To better frame our expectations, we use the {\tt StarTrack} population synthesis 
models \citep{Belczynski2002,Belczynski2008}.  We have employed the latest
version of  {\tt StarTrack}, which includes number of updates and revisions outlined below.
The initial distributions of binary parameters are consistent with observations
of massive O/B stars presented by~\cite{Sana2012}, with some minor modifications
discussed in~\cite{deMink2015}. The IMF is a 3 component broken power law
with boundaries at $M_{\rm ZAMS} = 0.08$, $0.5$, $1.0$, and $150 M_\odot$ and
slopes  of $-1.3$, $-2.2$ and $-2.3$ for the three mass ranges \citep{Kroupa2003}.
We assume a flat mass ratio distribution with $P(q)$
constant over the mass ratio range $0.1< q =M_2/M_1 < 1$ (e.g., \citealt{Kobulnicky2007}),
a binary orbital period distribution $P(\log p)\propto (\log p)^{-0.5}$
over the period range from $10^{0.15}$ to $10^{5.5}$~days,
and a power-law distribution of binary eccentricities $P(e) \propto
e^{-0.42}$ for $0<e<0.9$.
We then evolve single stars and primaries with masses from $5$--$150M_\odot$,
and secondaries with masses from $0.08$--$150M_\odot$ assuming a constant star
formation rate and Solar metallicity ($Z=Z_\odot=0.02$).

The evolutionary model adopted in our calculation corresponds to model M10
of \cite{Belczynski2017}. The improvements relevant for
massive star evolution include updates to the treatment of CE evolution
~\citep{Dominik2012}, the compact object masses produced by core collapse/supernovae
\citep{Fryer2012,Belczynski2012}, and the effects of pair-instability pulsation
supernovae and pair-instability supernovae \citep{Belczynski2016}.

In addition to the \cite{Blaauw1961} effects of symmetric mass loss, 
including neutrinos,
each NS and BH at formation is assigned a randomly oriented natal kick
velocity of magnitude
\begin{equation}
\label{eq:appd_bh_kick}
 V_{\rm NK} = (1 - f_{\rm fb}) \sqrt{V_x^2 + V_y^2 + V_z^2} \; \; ,
\end{equation}
where $V_x$, $V_y$, $V_z$ are three velocity components drawn from a
Maxwellian distribution with $\sigma = 265$~km/s \citep{Hobbs2005}. The
fallback parameter $f_{\rm fb}$ describes the fraction of the
stellar envelope that falls back onto the proto-compact object.
Specifically, the mass of the NS or BH is calculated from
\begin{equation}
\label{eq:appd_bh_mass}
 M_{\rm BH/NS} = 0.9 \; [M_{\rm proto} + f_{\rm fb} (M - M_{\rm proto})],
\end{equation}
where $M_{\rm proto} = 1.0M_\odot$ is the initial compact object mass formed
in core-collapse, $M$ is the pre-supernova mass of the star and the factor of
$0.9$ allows for $10\%$ of baryonic mass loss in neutrino emission.
The fallback parameter is estimated from the ``rapid'' supernova mechanism that
is able to reproduce the observed mass-gap between NSs and BHs from
\cite{Fryer2012}.\footnote{The formula for the model coefficient 
${\rm a_1}$ in Eqn.~16 of \cite{Fryer2012}
should read ${\rm a_1} = 0.25 - 1.275/(M - M_{\rm proto})$.}

There are two notable exceptions to this scheme for the natal kicks.
First, for the most massive BHs ($M_{\rm BH} \gtorder 10-15 M_\odot$),
there is no natal kick, as there is no mass loss ($f_{\rm fb}=1.0$).
Second, for NSs formed in electron capture supernova (ECS; \citealt{Miyaji1980}),
we assume no natal kick, as core-collapse is predicted to be very rapid and
the asymmetries needed to drive natal kicks 
may not develop \citep{Dessart2006, Jones2013,Schwab2015}.
The effects of neutrino mass loss are still included for both cases,
as is symmetric mass loss for the electron capture SNe.

Based on these models, 
Table~\ref{tab:models} summarizes the supernovae from systems that 
started as binaries.  Roughly 34\% of the ccSN associated
with binaries are the explosions of merger remnants, quite similar to the
$22_{-9}^{+26}\%$ found by \cite{Renzo2018}.  For those that are binaries
at the time of explosion, we divide the sample by the remnant formed
in the explosion (NS or BH) and the type of companion at the time
of the explosion: main sequence (MS), evolved/stripped (EV), 
white dwarf (WD), NS or BH.  We present the totals and then subdivide them
by whether the binary is bound or unbound after the explosion.  
There is a very small fraction ($\sim 0.1\%$) where the remnant
collides with the stellar companion after the explosion which
we will not track.  In the discussion that follows, we 
consider only the binaries in which the companion is stellar 
and ignore those in which the companion at the time of explosion
is a compact object.  We will also generally not distinguish 
between MS and evolved/stripped stellar companions.

Whether a stellar companion shows any peculiarities after the ccSN
will strongly depend on the separation at the time of the explosion
relative to the radius of the star.  
Figure~\ref{fig:edist} shows the distribution of the stellar binaries in the
ratio $a/R_*$ of the pre-SN semimajor axis $a$ to the stellar radius
$R_*$.  We show the four cases corresponding to binaries which become
unbound or remain bound and either NS or BH formation. 
We are only interested in cases where this
ratio is small, and these orbits have generally circularized and
we can ignore ellipticity. Geometrically,  
the star subtends fraction $R_*^2/4 a^2$ of the explosion, and
even for $a/R_*=3$, only 2.8\% of the ccSN energy is intercepted by
the companion.  In our models, only the explosions forming neutron 
stars with binaries that remain bound have even a small chance of
being significantly impacted by the explosion, and even then it is
only $\sim 1\%$ of these systems. \cite{Liu2015} estimate that
5\% have $a/R_* <5$, which is roughly consistent with Figure~\ref{fig:edist}.  
Hence, strongly shock-impacted companions are rare.

Figure~\ref{fig:mdist} shows the masses of stellar companions, again
divided by the type of remnant formed and whether the binary remains
bound.  As expected from the simple non-interacting model and also
found by \cite{Renzo2018}, the typical companion is quite massive,
with $\sim 88\%$ of the companions associated with the formation of
an NS having $M_* > 3M_\odot$.  Companions to newly formed black 
holes tend to be somewhat more massive.  

Figure~\ref{fig:vdist} shows the post-SN transverse velocities of the
systems with the same divisions by outcome.  For the disrupted systems,
the velocities of both the star and the remnant are shown.  
These are velocities relative to the pre-SN binary center of mass and 
would be applicable for motions relative to the center of the SNR or 
the relative motions of unbound systems.  Like
\cite{Renzo2018}, we find that the stars 
move surprisingly slowly. For the NS case, 90\% of the bound (unbound)
stars moving more slowly than $23$~km/s ($20$~km/s).  For the BH case
the velocities are still lower, with 90\% of the bound (unbound)
stars moving more slowly than $1$~km/s ($8$~km/s).  While the 
typical binary moves more slowly than the typical unbound star,
there is a high velocity tail to the binary distribution.

In \S\ref{sec:sample} we measure transverse velocities relative
to the rest frame defined by nearby stars.  These transverse
velocities will include the pre-SN motion of the binary relative
to nearby stars.  The typical one-dimensional velocity dispersion
of O stars is $10$~km/s (see, e.g., \citealt{Binney1998}), so
we add Gaussian deviates with this amplitude to the velocity
vectors from the BPS models.  This leads to the velocity 
distribution shown in Figure~\ref{fig:vdist2}. It is now
quite difficult to discern any differences in the stellar
velocities by binary status, although the systems associated
with forming a BH still have markedly lower velocities.  For
the bound and unbound NS (BH) systems, 90\% of the velocities
are less than $32$~km/s ($22$~km/s) and $30$~km/s ($23$~km/s), 
respectively.  
We should note that \cite{Renzo2018} discuss the statistics of
runaway stars unassociated with SNRs 
using the pre-SN binary frame (Figure~\ref{fig:vdist})
when they should be using velocities in the local stellar rest
frame (Figure~\ref{fig:vdist2}).

Finally, Figure~\ref{fig:bperiod} shows the period distribution of
the surviving star plus NS or BH binaries.  Using the standard
approximations for the size of the Roche lobe (e.g., \citealt{Paczynski1971}),
we also show the period distribution of the binaries
that have pericenters with Roche lobes smaller than the stellar
radius.   These are the systems which can be mass transfer binaries
immediately after the explosion.  They are relatively rare, consisting
of 3.6\% of the NS binaries and 0.1\% of the BH binaries.  These
could be underestimates if the impact of the explosion 
or a sudden increase in the tidal forces on the star drive
it to expand on a short time scale.
  
\begin{table}
  \centering
  \caption{ccSN From Initial Binaries By Outcome}
  \begin{tabular}{lccc}
  \hline
  \multicolumn{1}{c}{Outcome} &
  \multicolumn{1}{c}{Total} &
  \multicolumn{1}{c}{Single} &
  \multicolumn{1}{c}{Binary} \\
  \hline
Merger  & 0.338 &  $-$  & $-$ \\ 
NS$+$MS & 0.230 & 0.200 & 0.029 \\ 
NS$+$EV & 0.014 & 0.010 & 0.003 \\ 
NS$+$WD & 0.041 & 0.028 & 0.013 \\ 
NS$+$NS & 0.177 & 0.171 & 0.006 \\ 
NS$+$BH & 0.026 & 0.025 & 0.001 \\ 
BH$+$MS & 0.044 & 0.021 & 0.024 \\ 
BH$+$EV & 0.005 & 0.002 & 0.002 \\ 
BH$+$WD & 0.000 & 0.000 & 0.000 \\ 
BH$+$NS & 0.075 & 0.075 & 0.000 \\ 
BH$+$BH & 0.049 & 0.039 & 0.010 \\ 
  \hline
  \multicolumn{4}{l}
   {For the explosions in binaries, the first } \\
  \multicolumn{4}{l}
   {entry is the compact object being formed } \\
  \multicolumn{4}{l}
   {and the second entry is the binary companion} \\ 
  \multicolumn{4}{l}
    {at the time. } \\
   \end{tabular}
\label{tab:models}
\end{table}

\begin{figure}
\centering
\includegraphics[width=0.45\textwidth]{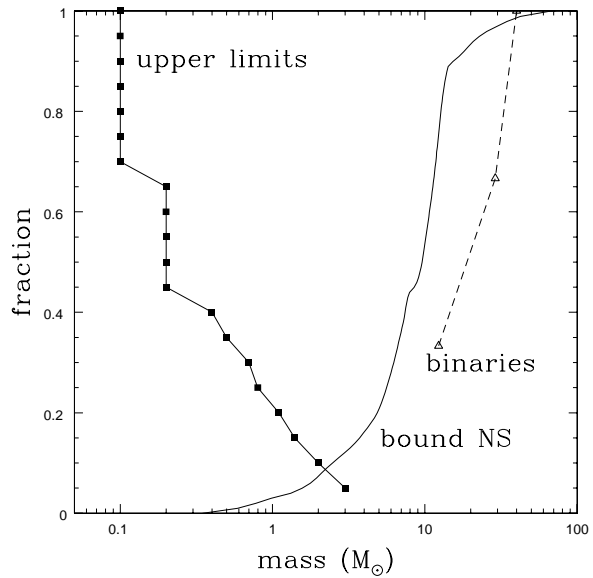}
\caption{
  The integral distributions of the three binary masses (dashed, triangles) and the
  20 upper limits (solid, squares) for the SNRs with clearly associated compact
  remnants.  For comparison,
  the mass distribution of bound NS binaries in our BPS models from 
  \S\protect\ref{sec:theory} is also shown.   
  }
\label{fig:masses}
\end{figure}

\begin{figure}
\centering
\includegraphics[width=0.45\textwidth]{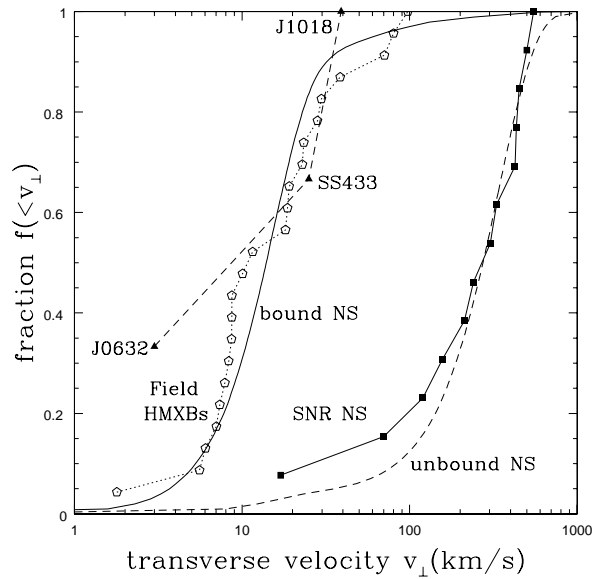}
\caption{
  The integral distributions of the three binary transverse velocities (dashed, triangles) 
  and the non-binary NS transverse velocities (solid, squares). For comparison,
  we also show the transverse velocities of field HMXBs from the catalog
  \protect\cite{Liu2006}. Since HMXBs are short lived, their velocities are
  little affected by their post-SN Galactic orbit and should be directly comparable 
  to the binaries in the SNRs.  The transverse velocity distributions of bound star/NS
  binaries and unbound NS from our BPS models in \S\protect\ref{sec:theory} are also  
  shown.  These include an additional velocity dispersion of $\sigma = 10$~km/s for 
  each velocity component because we are measuring the transverse velocities relative
  to the local stars (i.e., Figure~\protect\ref{fig:vdist2} rather than Figure~\protect\ref{fig:vdist}).  
  }
\label{fig:velos}
\end{figure}

\def\gray{$\gamma$-ray}
\def\sp{\hphantom{-}}
\begin{table*}
   \centering
   \caption{Limits on Companions}
   \begin{tabular}{rrrrrrrrr}
   \hline
     \multicolumn{1}{c}{SNR}
    &\multicolumn{1}{c}{Compact}
    &\multicolumn{1}{c}{Discovery}
    &\multicolumn{1}{c}{$d$}
    &\multicolumn{1}{c}{$E(B-V)$}
    &\multicolumn{1}{c}{Limit}
    &\multicolumn{1}{c}{Mass} 
    &\multicolumn{1}{c}{$v_\perp$} 
    &\multicolumn{1}{c}{Known} \\
    &\multicolumn{1}{c}{Source}
    &
    &\multicolumn{1}{c}{(kpc)}
    &\multicolumn{1}{c}{(mag)}
    &\multicolumn{1}{c}{(mag)}
    &\multicolumn{1}{c}{($M_\odot$)}
    &\multicolumn{1}{c}{(km/s)}
    &\multicolumn{1}{c}{Binary} \\
   \hline
   G034.7$-$00.4 &J1856$+$0113            &radio   &$3.0\pm0.5$   &$2.2\pm0.7$   &$H>19.0$  &$<0.7$         &$\cdots$          \\
   G039.7$-$02.0 &SS~433                  &radio   &$5.0\pm0.7$   &$1.9\pm0.9$   &$\cdots$  &$12.3\pm 3.3$  &$25\pm 8$    &Yes \\
   G065.7$+$01.2 &J1952.2$+$2925          &X-ray   &$3.0\pm2.0$   &$0.7\pm0.5$   &$z>18.2$  &$<3.0$         &$\cdots$    \\
   G069.0$+$02.7 &PSR~B1951+32            &radio   &$2.0\pm0.5$   &$0.5\pm0.2$   &$V>24.3$  &$<0.5$         &$240\pm40$    \\
   G078.2$+$02.1 &PSR~J2021$+$4026        &\gray   &$1.5\pm0.5$   &$1.3\pm0.3$   &$H>19.0$  &$<0.2$         &$\sim 550$    \\
   G106.3$+$02.7 &PSR~J2229$+$6114        &radio   &$3.0\pm1.0$   &$1.1\pm0.2$   &$R>23$    &$<0.8$         &$\cdots$    \\
   G109.1$-$01.0 &AXP 1E~2259$+$586       &X-ray   &$4.1\pm0.7$   &$1.3\pm0.4$   &$K>21.7$  &$<0.1$         &$157\pm17$   \\
   G111.7$-$02.1 &Cas~A                   &X-ray   &$3.4\pm0.3$   &$1.7\pm0.9$   &$J>26.2$  &$<0.1$         &$\sim 330$    \\
   G114.3$+$00.3 &PSR~B2334$+$61          &radio   &$3.2\pm1.7$   &$1.0\pm0.7$   &$z>22.3$  &$<1.1$         &$212\pm268$    \\
   G119.5$+$10.2 &PSR J0007+7303          &X-ray   &$1.4\pm0.3$   &$0.5\pm0.1$   &$r>27.6$  &$<0.1$         &$\sim 450$    \\
   G130.7$+$03.1 &PSR~J0205$+$6449        &X-ray   &$4.5\pm1.6$   &$0.7\pm0.1$   &$i>25.5$  &$<0.2$         &$17\pm9$    \\
   G180.0$-$01.7 &PSR~J0538$+$2817        &radio   &$1.5\pm0.4$   &$0.5\pm0.1$   &$z>22.3$  &$<0.2$         &$425 \pm 94$    \\
   G184.6$-$05.8 &Crab                    &radio   &$2.5\pm0.5$   &$0.5\pm0.1$   &$J>14.8$  &$<2.0$         &$\sim 120$    \\
   G189.1$+$03.0 &CXO~J061705.3$+$222127  &X-ray   &$1.7\pm0.3$   &$1.2\pm0.1$   &$z>22.3$  &$<0.2$         &$\sim 500$    \\
   G205.5$+$00.5 &HESS~J0632$+$057        &\gray   &$1.8\pm0.2$   &$0.6\pm0.1$   &fit       &$20$-$36$      &$3\pm1$   &Yes  \\
   G260.4$-$03.4 &PSR~J0821$-$4300        &X-ray   &$1.3\pm0.3$   &$0.5\pm0.1$   &$I>25.6$  &$<0.1$         &$433\pm 126$    \\
   G263.9$-$03.3 &Vela                    &radio   &$0.29\pm0.02$ &$0.1\pm0.1$   &$R>23.9$  &$<0.1$         &$70 \pm 8$   \\ 
   G266.2$-$01.2 &CXOU~J085201.4$-$461753 &X-ray   &$0.7\pm0.2$   &$0.5\pm0.2$   &$R>25.6$  &$<0.1$         &$\cdots$    \\
   G284.3$-$01.8 &1FGL~J1018.6-5856       &\gray   &$5.4\pm3.0$   &$1.5\pm0.2$   &fit       &$29$-$61$      &$39\pm 7$  &Yes \\
   G291.0$-$00.1 &CXOU~J111148.6$-$603926 &X-ray   &$5.0\pm2.0$   &$1.1\pm0.1$   &$G>22$    &$<1.4$         &$303\pm130$    \\
   G296.5$+$10.0 &PSR~J1210$-$5226        &X-ray   &$2.1\pm1.8$   &$0.2\pm0.1$   &$R>27.1$  &$<0.2$         &$\cdots$   \\
   G320.4$-$01.2 &PSR~J1513$-$5908        &X-ray   &$5.2\pm1.4$   &$1.5\pm0.2$   &$H>20.6$  &$<0.4$         &$\cdots$    \\
   G332.4$-$00.4 &1E~161348$-$5055        &X-ray   &$3.1\pm0.5$   &$1.7\pm0.7$   &$K_s>22.1$&$<0.1$         &$\cdots$             \\
% N(H) too high
%   G049.2$-$00.7 &CXO~J192318.5$+$1403035 &X-ray   &$4.9\pm0.6$   &$3.1\pm0.6$   &$H>19.0$  &$<1.2$         &$\sim 360$    \\
%   G076.9$+$01.0 &PSR~J2022$+$3842        &X-ray  &$10.0\pm2.0$   &$2.9\pm0.5$   &$H>19.0$  &$<2.6$         &$\cdots$    \\
%   G292.2$-$00.5 &PSR~J1119$-$6127        &X-ray   &$8.4\pm0.4$   &$4.3\pm1.0$   &$K>22$    &$<0.3$         &$<500$    \\
% Distance too high
%   G292.0$+$01.8 &PSR~J1124$-$5916        &radio   &$6.2\pm0.9$   &$0.5\pm0.1$   &$K>22.8$  &$<0.1$         &$440$    \\
% Type Ia
%   G315.4$-$02.3 &$[$GV2003$]$N           &X-ray   &$25.\pm0.5$   &$1.0\pm0.1$   &fit       &$0.9$-$1.2$    &           & Yes\\
   \hline
   \multicolumn{9}{l}{The Type column indicates the discovery method.  The distance, $d$, and extinction, $E(B-V)$,
                     columns are rough } \\
   \multicolumn{9}{l}{summaries based on the discussion of each object in Appendix~\ref{sec:accepted}.
                     As discussed in \S\ref{sec:sample}, the results are robust to } \\
   \multicolumn{9}{l}{reasonable changes in these
                     estimates.  The Mag column gives the most constraining magnitude limit on the presence of a} \\
   \multicolumn{9}{l}{stellar companion, excluding the three known binaries.  The Mass column is the upper
                     mass limit implied by the magnitude } \\
   \multicolumn{9}{l}{limit or a mass estimate for the stellar companion in
                     the binary.  For SS~433 we use the mass from \cite{Hillwig2008}. } \\
   \multicolumn{9}{l}{The transverse velocity measurements or  limits are given in the $v_\perp$ } \\
   \multicolumn{9}{l}{column, and the Known Binary column flags the known binaries.} \\ 
   \end{tabular}
   \label{tab:results}
\end{table*}

\begin{table*}
   \centering
   \caption{HMXB Transverse Velocities}
   \begin{tabular}{lrrrrrrrr}
   \hline
     \multicolumn{1}{c}{HMXB}
    &\multicolumn{1}{c}{Gaia ID}
    &\multicolumn{1}{c}{$v_\perp$}
    &\multicolumn{1}{c}{$\pi$}
    &\multicolumn{1}{c}{$PM RA$}
    &\multicolumn{1}{c}{$PM Dec$}
    &\multicolumn{1}{c}{$N_*$}
    &\multicolumn{1}{c}{$\langle PM RA\rangle$}
    &\multicolumn{1}{c}{$\langle PM Dec\rangle$} \\
    &
    &\multicolumn{1}{c}{(km/s)}
    &\multicolumn{1}{c}{(mas)}
    &\multicolumn{1}{c}{(mas/year)}
    &\multicolumn{1}{c}{(mas/year)}
    &
    &\multicolumn{1}{c}{(mas/year)}
    &\multicolumn{1}{c}{(mas/year)} \\
   \hline
   1H~1253$-$761 &5837600152935767680 &$ 19\pm   1$ &$  4.70\pm  0.03$ &$-27.18\pm  0.06$ &$ -9.02\pm  0.05$ &1499 &$-10.42\pm 13.62$ &$ -0.40\pm  7.43$ \\ 
1H~1249$-$637 &6055103928246312960 &$ 10\pm   1$ &$  2.38\pm  0.12$ &$-12.51\pm  0.16$ &$ -3.98\pm  0.15$ &1737 &$ -8.72\pm  7.94$ &$ -0.67\pm  4.66$ \\ 
2S~1145$-$619 &5334823859608495104 &$  6\pm   1$ &$  0.42\pm  0.04$ &$ -6.22\pm  0.05$ &$  1.46\pm  0.05$ &1948 &$ -6.71\pm  2.61$ &$  1.24\pm  1.73$ \\ 
4U~1258$-$61 &5863533199843070208 &$ 28\pm   3$ &$  0.47\pm  0.03$ &$ -4.23\pm  0.04$ &$ -0.32\pm  0.05$ &260 &$ -6.98\pm  3.21$ &$ -0.77\pm  1.90$ \\ 
1H~1255$-$567 &6060547331128876928 &$ 11\pm   1$ &$  8.95\pm  0.23$ &$-28.15\pm  0.22$ &$-10.34\pm  0.34$ &1525 &$ -9.11\pm 18.94$ &$  0.11\pm 12.31$ \\ 
1H~1555$-$552 &5884544931471259136 &$  7\pm   0$ &$  0.72\pm  0.04$ &$ -3.11\pm  0.06$ &$ -3.29\pm  0.05$ &1359 &$ -3.86\pm  3.49$ &$ -4.04\pm  2.83$ \\ 
1H~0739$-$529 &5489434710755238400 &$  8\pm   1$ &$  1.53\pm  0.04$ &$ -4.53\pm  0.09$ &$  8.60\pm  0.09$ &1324 &$ -3.22\pm  6.42$ &$  6.15\pm  8.66$ \\ 
1WGA~J0648.0$-$4419 &5562023884304074240 &$  8\pm   1$ &$  1.97\pm  0.06$ &$ -4.11\pm  0.11$ &$  5.67\pm  0.12$ &1875 &$ -0.60\pm  8.00$ &$  4.94\pm 13.37$ \\ 
4U~0900$-$40 &620657678322625920 &$ 70\pm   7$ &$  0.38\pm  0.03$ &$ -4.96\pm  0.05$ &$  9.09\pm  0.05$ &1470 &$ -4.09\pm  2.37$ &$  3.46\pm  2.54$ \\ 
4U~1700$-$37 &5976382915813535232 &$ 80\pm   9$ &$  0.55\pm  0.06$ &$  2.22\pm  0.09$ &$  4.95\pm  0.07$ &1705 &$ -2.02\pm  3.12$ &$ -3.25\pm  3.00$ \\ 
RX~J1744.7$-$2713 &4060784345959549184 &$  8\pm   1$ &$  0.83\pm  0.06$ &$ -0.96\pm  0.10$ &$ -2.06\pm  0.08$ &1301 &$ -1.88\pm  3.92$ &$ -3.26\pm  3.81$ \\ 
IGR~J17544$-$2619 &4063908810076415872 &$ 22\pm   3$ &$  0.35\pm  0.05$ &$ -0.65\pm  0.08$ &$ -0.53\pm  0.07$ &1827 &$ -0.53\pm  2.30$ &$ -2.22\pm  2.58$ \\ 
RX~J1826.2$-$1450 &4104196427943626624 &$ 97\pm  10$ &$  0.48\pm  0.05$ &$  7.43\pm  0.09$ &$ -8.00\pm  0.07$ &1514 &$ -0.99\pm  2.33$ &$ -2.77\pm  2.81$ \\ 
XTE~J1901$+$014 &4268294763117802368 &$ 38\pm   4$ &$  0.64\pm  0.08$ &$ -4.07\pm  0.11$ &$ -7.53\pm  0.10$ &1737 &$ -0.88\pm  3.16$ &$ -3.49\pm  3.97$ \\ 
1A~0535$+$262 &3441207615229815040 &$ 18\pm   2$ &$  0.44\pm  0.05$ &$ -0.63\pm  0.09$ &$ -3.04\pm  0.07$ &1361 &$  0.92\pm  1.90$ &$ -2.27\pm  2.42$ \\ 
1H~0556$+$286 &3431561565357225088 &$  8\pm   1$ &$  0.39\pm  0.06$ &$  0.61\pm  0.11$ &$ -2.72\pm  0.09$ &1553 &$  0.72\pm  1.78$ &$ -2.05\pm  2.15$ \\ 
4U~0352$+$309 &168450545792009600 &$ 18\pm   1$ &$  1.23\pm  0.06$ &$ -1.40\pm  0.10$ &$ -2.25\pm  0.07$ &1963 &$  2.67\pm  6.08$ &$ -4.63\pm  5.79$ \\ 
4U~1956$+$35 &2059383668236814720 &$ 23\pm   2$ &$  0.42\pm  0.03$ &$ -3.88\pm  0.05$ &$ -6.17\pm  0.05$ &1305 &$ -2.41\pm  2.16$ &$ -4.72\pm  2.84$ \\ 
EXO~051910$+$3737.7 &84497471323752064 &$  7\pm   1$ &$  0.75\pm  0.06$ &$  1.44\pm  0.12$ &$ -4.12\pm  0.07$ &1870 &$  1.32\pm  2.79$ &$ -2.87\pm  3.31$ \\ 
RX~J2030.5$+$4751 &2083644392294059520 &$  7\pm   1$ &$  0.37\pm  0.03$ &$ -3.05\pm  0.05$ &$ -4.62\pm  0.05$ &1388 &$ -2.61\pm  2.34$ &$ -4.24\pm  2.76$ \\ 
1H~2202$+$501 &1979911002134040960 &$ 29\pm   3$ &$  0.84\pm  0.04$ &$  2.40\pm  0.07$ &$ -0.27\pm  0.07$ &1112 &$ -2.09\pm  4.25$ &$ -2.97\pm  3.68$ \\ 
1E~0236.6$+$6100 &465645515129855872 &$  1\pm   0$ &$  0.38\pm  0.04$ &$ -0.30\pm  0.04$ &$ -0.08\pm  0.07$ &1903 &$ -0.22\pm  1.89$ &$ -0.20\pm  1.87$ \\ 
RX~J0146.9$+$6121 &511220031584305536 &$  5\pm   0$ &$  0.37\pm  0.03$ &$ -0.88\pm  0.03$ &$  0.03\pm  0.04$ &1599 &$ -0.92\pm  1.70$ &$ -0.40\pm  1.47$ \\ 

   \hline
   \multicolumn{9}{l}{For each HMXB we give the associated Gaia ID, the final estimate of the transverse
    velocity $v_\perp$, the Gaia parallax ($\pi$), the proper motions in RA } \\
   \multicolumn{9}{l}{($PM RA$) and Dec ($PM Dec$), the
    number of stars $N_*$ used to define the local standard of rest, and the mean and dispersion of
    the proper } \\
    \multicolumn{9}{l}{motions of these stars in RA ($\langle PM RA\rangle$) and Dec ($\langle PM Dec\rangle$).
    The uncertainty in the mean proper motion is smaller than the dispersion by } \\
   \multicolumn{9}{l}{$(N_*-1)^{-1/2}$.} \\
   \end{tabular}
   \label{tab:hmxb}
\end{table*}

%\begin{table*}
%   \centering
%   \caption{LMXB Transverse Velocities}
%   \begin{tabular}{lrrrrrrrr}
%   \hline
%     \multicolumn{1}{c}{LMXB}
%    &\multicolumn{1}{c}{Gaia ID}
%    &\multicolumn{1}{c}{$v_\perp$}
%    &\multicolumn{1}{c}{$\mu$}
%    &\multicolumn{1}{c}{$PM RA$}
%    &\multicolumn{1}{c}{$PM Dec$}
%    &\multicolumn{1}{c}{$N_*$}
%    &\multicolumn{1}{c}{$\langle PM RA\rangle$}
%    &\multicolumn{1}{c}{$\langle PM Dec\rangle$} \\
%    % 
%    &
%    &\multicolumn{1}{c}{(km/s)}
%    &\multicolumn{1}{c}{(mas)}
%    &\multicolumn{1}{c}{(mas/year)}
%    &\multicolumn{1}{c}{(mas/year)}
%    &
%    &\multicolumn{1}{c}{(mas/year)}
%    &\multicolumn{1}{c}{(mas/year)} \\
%   \hline
%   \input lmxbs/lmxb.tex
%   \hline
%   \multicolumn{9}{l}{The columns are the same as in Table~\ref{tab:hmxb}.} \\
%   \end{tabular}
%   \label{tab:lmxb}
%\end{table*}

\section{Sample}
\label{sec:sample}

To build our sample of SNRs, we started with the Manitoba data base of SNRs (\citealt{Ferrand2012}).  In
practice, this data base is a mixture of SNRs and pulsar wind nebulae
(PWN), so we also required the SNR to be in the SNR catalog of 
\cite{Green2014}.  Next we required the SNR to have a reported
distance that could be closer than 5~kpc. 
Where possible, we required an X-ray estimate of the column density 
$N(H) < 10^{22}$~cm$^{-2}$.  If no X-ray estimate was available, we
required a total Galactic $A_V< 5$~mag based on \cite{Schlafly2011}.  We assume a relation
between $N(H)$ and extinction of $E(B-V)=1.7 N(H) \times 10^{-22}$
(\citealt{Bohlin1978}).  For $R_V=3.1$, $N(H) = 10^{22}$~cm$^{-2}$
corresponds to $A_V=5.3$~mag or $E(B-V)=1.7$.  These criteria led
to an initial list of 54 SNRs.

We then investigated the individual SNRs in detail and divided them into three
categories. First, there are 23 SNRs with clearly associated
compact objects identified as either radio pulsars, X-ray
sources or $\gamma$-ray sources.  Second, there are 26 SNRs without 
clearly associated compact objects.  This includes cases
with X-ray fluxes below $10^{-13}$~erg~cm$^{-2}$~s$^{-1}$
where there are generally multiple candidates for an associated
compact object, but too few counts to determine their nature.   
Finally, there are 5 SNRs which are dropped.  In four cases, they are
SNRs associated with Type~Ia SNe (Kepler, Tycho, G315.4$-$02.3
and G327.6$+$14.6), and in one case (G011.1$+$00.1) the 
distance in the data base is for a NS in the foreground
of a significantly more distant SNR.  
Appendix~\ref{sec:accepted} summarizes the  properties of the
SNRs with clearly associated compact objects, Appendix~\ref{sec:nocompact} summarizes 
those without, and Appendix~\ref{sec:rejected} summarizes the
rejected SNRs.  Of the sample with compact objects, 8, 12
and 3 were {\it first} identified as radio, X-ray and $\gamma$-ray sources, respectively.
Three of these sources, SS~433 (see the review by \citealt{Margon1984}), 
HESS~J0632$+$057 (\citealt{Hinton2009}) and 1FGL~J1018.6-5856 (\citealt{Corbet2011}),
are known binaries.  

For each of the accepted SNRs we searched for prior optical or near-IR
studies of the compact object.  If none were available, we searched for 
the source in other catalogs, in particular, PanSTARRS (\citealt{Chambers2016}),
2MASS (\citealt{Skrutskie2006}), UKIDSS (\citealt{Lucas2008}), 
APASS (\citealt{Henden2016})  and 
Gaia DR2 (\citealt{Gaia2018}).  We adopted distance estimates from 
the literature, supplemented by the Gaia parallaxes of the known
binaries.  For many of the sources we could check the
X-ray estimate of the extinction based on the hydrogen
column density $N(H)$ against the PanSTARRS 3D dust distribution models of 
\cite{Green2015}.  The two extinction estimates generally agreed
reasonably well.  These distance and extinction estimates were 
converted into the rough priors given in Table~\ref{tab:results}. 
We assume minimum extinction uncertainties of $0.1$~mag.
To interpret the
magnitudes, we used Solar metallicity PARSEC v1.2S (\citealt{Chen2015})
isochrones sampled in $\log(\hbox{age/years})$ from $6.0$ to $7.5$ increments of
0.01~dex.  Note that at these young ages, the lower mass stars are not on
the main sequence when the primary explodes -- they are significantly
more luminous pre-main sequence stars.

For the sources without stellar companions, we use the optical and 
near-IR fluxes or flux limits to set upper bounds on the mass of 
any stellar companion.  Some of the systems (e.g., the Crab, see
Appendix~\ref{sec:accepted})
have optical/near-IR counterparts due to emission from the NS
and we simply use these fluxes as upper limits. 
Given an apparent magnitude $m_i$ and a model absolute magnitude $M_i$
for filter $i$, we have that $m_i = M_i + \mu + R_i E(B-V) $ where
we use the PARSEC estimates of $R_i$ for each filter based on
a \cite{Cardelli1989} $R_V=3.1$ extinction curve.  Given the 
distance (modulus) prior ($\mu_0 \pm \sigma_\mu$), the extinction
prior ($E_0\pm \sigma_E$) and the relation between apparent and
absolute magnitudes, we optimize the goodness of fit statistic
$$
  \chi^2 = \left( { \mu - \mu_0 \over \sigma_\mu } \right)^2
    + \left( { E - E_0 \over \sigma_E } \right)^2
$$
to estimate the distance modulus $\mu$ and extinction $E$ for each model
star on the isochrones.  Because the fits are under-constrained,
essentially having two constraints (on $\mu$ and $E$) for four
variables ($\mu$, $E$, temperature $T_*$ and luminosity $L_*$), there are always solutions
with $\chi^2=0$.   As a conservative bound on the mass, we find the
maximum mass for each filter which satisfies $\chi^2<4$, and then
report in Table~\ref{tab:results} the magnitude limit for the most 
constraining filter and the associated maximum mass.  Generally, 
other similar filters (e.g., $J$ and $H$ if the best limit came 
from $K$) give similar but slightly weaker constraints.

The constraints on distance and extinction in Table~\ref{tab:results}
are designed to roughly cover the range of values and uncertainties
found in our survey of the literature for each source.  Despite the
somewhat uncertain nature of the estimates for many sources, our results 
are robust to reasonable changes in these estimates for two reasons.
First, we are not particularly interested in the difference between a mass
limit of (for example) $0.2 M_\odot$ and $0.4 M_\odot$.  As outlined
in the introduction and discussed further in \S\ref{sec:theory}, all
that really matters is that the source cannot have a (say) $3 M_\odot$
stellar counterpart since this already encompasses a large fraction
of the expected secondary population.  Second, particularly for this level of 
accuracy, the mass estimates are robust to changes in distance
or extinction because luminosity is a 
steep function of mass for (pre-)main sequence stars, $L \propto M^x$ 
with $ x\simeq 2$-$3$.  Since $L$ depends on distance
as $d^2$, the mass depends on distance as $M \propto d^{2/x}$ 
and even a factor of two change in distance changes the mass by
only 60\% for $x=3$.  Similarly, $L$ depends on extinction as $10^{R_i E/2.5}$,
so the mass depends on extinction as $10^{R_i E / 2.5 x}$.  For 
$x=3$, changing the extinction $E(B-V)$ by 1 magnitude, changes the mass by a factor of
$\sim 2.2$ at R band (effectively the bluest filter used in
Table~\ref{tab:results}), but only by 30\% at J band and 11\% at K. 

A different procedure is required for the actual binaries. For
HESS~J0632$+$057 and 1FGL~J1018.6-5856 we fit the spectral 
energy distributions (SED) using the PARSEC models, adding 
a spectroscopic temperature prior and a Gaia parallax prior
(see Appendix~\ref{sec:accepted}). 
We did this by fitting the photometry with the priors on 
distance and extinction given in Table~\ref{tab:results} and then adding
terms to the $\chi^2$ for parallax ($\pi \pm \sigma_\pi$)
and temperature ($T \pm \sigma_T$).  This treatment of the
distance and the parallax is imperfect, but there is no need
to do a more complex non-linear fit for our purposes.  For
HESS~J0632$+$057 we get a mass of $29 M_\odot$ 
($20 M_\odot < M < 36 M_\odot$), somewhat larger than 
the estimate of $13 M_\odot < M < 19M_\odot$ by 
\cite{Aragona2010}. The difference is likely that
\cite{Aragona2010} correct for a significant amount
of disk emission, while we did not.
For 1FGL~J1018.6-5856, we get $40M_\odot$
($29M_\odot < M < 61M_\odot$) while \cite{Napoli2011}
find $31 M_\odot$.  The optical and near-IR emission of
SS~433 is completely dominated by accretion, and there 
is some debate over the mass of the star.  For our
study, we simply 
adopt the estimate by \cite{Hillwig2008} of 
$(12.3 \pm 3.3)M_\odot$.  
The mass limits and binary masses are shown
in Figure~\ref{fig:masses}.

Where possible, we also include estimates of the transverse
velocity $v_\perp$ in Table~\ref{tab:results}.  Many come from
the offset of the source from the geometric or expansion
center of the SNR, while others are from actual proper
motion measurements (see Appendix~\ref{sec:accepted}).  
\cite{Holland2017} has a good
discussion of the reliability of these estimates. For
our purposes, we are interested in the order of magnitude
of the velocities rather than their precise values.  
Where there are proper motion measurements, we used
Gaia DR2 proper motions of stars near the estimated
distance of the compact object to empirically determine
the local standard of rest without any reference to
a kinematic model for the Galaxy.

In particular, for the three binaries, we compare the
Gaia DR2 proper motion of each binary to stars whose
parallax is within $1\sigma$ of the binary's parallax.
We used a search region (5 to 30~arcmin) centered on the 
source that was large enough to include 1000 to 2000 stars 
meeting the parallax criterion.  We then computed the
proper motion of the stars and the dispersion of the
proper motions around the mean.  The transverse velocity 
is then the motion of the binary minus
the local mean multiplied by the distance determined
from the parallax, with an uncertainty determined by
the standard propagation of errors.  If
we use the same procedure for the pulsars with VLBI parallaxes,
we find results consistent with previously measured 
transverse velocities using Galactic kinematic models
to define the local standard of rest.  Table~\ref{tab:results}
and Figure~\ref{fig:velos} use our estimates of $v_\perp$.

Since all three binaries in our sample are HMXBs, we carried 
out a similar analysis of
the transverse velocities of HMXBs from the catalog
of \cite{Liu2006}.  The catalog contains 114 binaries,
67 of which have an optical magnitude.  Of these, we
found good matches in Gaia DR2 for 57 and we kept the 23
with $V<19$~mag, parallaxes larger than 
$0.333$~mas (i.e., $d<3$~kpc), and parallax errors
at least 5 times smaller than the parallax.  
As with the binaries in the SNRs, we then selected 
$N_*=1000$-$2000$ nearby stars with parallaxes within $1\sigma$ 
of the parallax of the binary to define the local standard 
of rest and defined the transverse velocity of the binary
by the difference between the proper motion of the binary
and the mean proper motion of the nearby stars. These 
estimates are given in Table~\ref{tab:hmxb}.

%We also estimated the transverse
%velocities of the LMXBs in \cite{Liu2007}.  For the LMXBs
%we required a Gaia magnitude $G<19$, a parallax larger
%than $0.333$~mas, and a parallax error at least 3 times
%smaller than the parallax.
%Even after allowing the larger fractional parallax
%errors we were left with only 6 transverse
%velocity estimates from an initial catalog of 49 systems.
%The resulting transverse velocities are given in 
%Tables~\ref{tab:hmxb} and \ref{tab:lmxb}, respectively,
%and shown in Figure~\ref{fig:velos}.
%The tables include the parallaxes and proper motions of
%the binaries along with the mean and dispersion of the
%parallaxes of the stars used to define the local standard
%of rest.  The uncertainty in the mean local proper motion
%is then smaller than the dispersion by $(N_*-1)^{-1/2}$.

\begin{table}
  \centering
  \caption{Estimated Binary Fractions $f$}
  \begin{tabular}{lccccl}
  \hline
  \multicolumn{1}{c}{Case} &
  \multicolumn{1}{c}{$N_b$} &
  \multicolumn{1}{c}{$N_s$} &
  \multicolumn{1}{c}{median} &
  \multicolumn{1}{c}{90\% conf} &
  \multicolumn{1}{c}{comments}
          \\
  \hline
Non-Interacting NS  &$0$ &$23$ &           &$<0.091$        & $<3M_\odot$\\
                    &$0$ &$22$ &           &$<0.095$        & $<2M_\odot$\\
                    &$0$ &$21$ &           &$<0.099$        & $<M_\odot$\\
Interacting BH      &$1$ &$22$ &$0.069$    &$0.012$-$0.18$  & no bias \\
                    &$1$ &$42$ &$0.038$    &$0.008$-$0.10$  & naive Ia \\
                    &$1$ &$48$ &$0.033$    &$0.006$-$0.09$  & full bias \\
Interacting NS      &$2$ &$21$ &$0.109$    &$0.035$-$0.24$  & no bias \\
                    &$2$ &$41$ &$0.060$    &$0.019$-$0.14$  & naive Ia \\
                    &$2$ &$47$ &$0.053$    &$0.017$-$0.12$  & full bias \\
Interacting BH/NS   &$3$ &$20$ &$0.151$    &$0.059$-$0.29$  & no bias \\
                    &$3$ &$40$ &$0.083$    &$0.032$-$0.17$  & naive Ia \\
                    &$3$ &$46$ &$0.082$    &$0.032$-$0.17$  & full bias \\
  \hline
   \end{tabular}
\label{tab:statistics}
\end{table}

\section{Results}
\label{sec:results}

Figure~\ref{fig:masses} shows the resulting distribution of 
binary masses and upper limits for our sample of 23 
SNRs with remnants.  All three binaries are HMXBs and we find no low mass 
companions. SS~433 is a Roche lobe accretion system, and the
other two are X-ray and $\gamma$-ray sources due to interactions
between the stellar and pulsar winds.   
If we have $N_b$ binaries and $N_s$ non-binaries, then
the probability distribution for the binary fraction $f$
is the binomial distribution
\begin{equation}
  { dP \over df } \propto f^{N_b} \left( 1 -f \right)^{N_s}.
  \label{eqn:binomial}
\end{equation}
We can make several choices for $N_b$.  First, the systems we
can find are non-interacting NS binaries and interacting
NS or BH binaries.  We cannot find non-interacting BH
binaries with our search procedures. 
Second, the $\gamma$-ray binaries are believed to be
NS systems, while SS~433 is likely a BH binary
(see, e.g., \citealt{Hillwig2008} for a discussion). The
case $N_b=0$ provides an upper limit on the fraction
of SNRs with non-interacting NS binaries, the
case $N_b=1$ constrains
the fraction with interacting BH binaries,  the case
$N_b=2$ constrains the fraction of SNR with interacting NS
binaries, and, finally, the case $N_b=3$ constrains
the fraction of SNRs with interacting binaries 

We also have to decide on the value of $N_s$.  This
matters in three contexts.  First, our upper mass limits
are not uniform, so if we consider lower mass limits
we have fewer SNRs where data with the necessary depth
exists.  Second, we looked at 49 remnants but 
were only able to clearly identify compact objects to check
for stellar companions in 23.  The latter point 
should only be an issue for the interacting binaries
since the failure to identify a compact remnant should
have no consequences for our selection of non-interacting
binaries. Third, some of these 49 remnants are probably
due to Type~Ia SNe.  Based
on local volumetric rates (e.g., comparing \citealt{Horiuchi2010}
and \citealt{Horiuchi2011}), the Type~Ia rate is
roughly 20\% of the ccSN rate, although there remain
significant uncertainties.  We are also looking at SNRs,
not SNe, and SNRs from different types of SNe will
have different lifetimes (e.g., \citealt{Sarbadhicary2017}). 
We will consider the consequences of
a ``naive'' Type~Ia correction by assuming that 20\%
of the remnants are due to Type~Ia SNe.  However,
we should take 20\% of 53 SNR because we already
rejected 4 SNR for being Type~Ia.  So roughly,
we expect $10$ Type~Ia remnants, 4 of which are
already recognized, leaving 6 hiding amongst the 49.  

For the non-interacting binaries, we need to select
a companion mass limit.  The simplest possibility is simply to
set a limit on the fraction with $M_* > 3M_\odot$, 
corresponding to the weakest mass limit in 
Table~\ref{tab:results}.  In our BPS models, 
$12\%$ of surviving companions are less massive 
than $3M_\odot$ (see Fig.~\ref{fig:mdist}).
If we want to make an estimate for lower masses,
we must drop remnants with weaker limits.  So for
$M_* > 2 M_\odot$ ($>1 M_\odot$) we must drop
$1$ ($2$) systems to have $N_s=22$ ($21$).  Here
we assume that the Crab and PSR~B2334$+$61
are not binaries independent of the mass limits in
Table~\ref{tab:results} because they are radio pulsars
without timing residuals interpreted as due to binarity 
(see Appendix~A).  In practice, the mass limit for the
Crab in Table~\ref{tab:results} could also be made much
stricter by subtracting a model for the non-thermal
emissions of the pulsar.
In the BPS models only 7\% (3\%) of bound companions
are below these mass limits.

Interacting binaries have enhanced high energy emission and 
are more detectable, so the sample of 23 SNRs with compact
objects likely contains all the interacting binaries in 
the full sample of 49 SNRs.  This
suggests that we should then use $N_s=49-N_b$ for the 
statistics of the interacting binary cases.
However, we also consider a ``naive'' Type~Ia
correction where we assume that 6 of the remaining
SNRs are unidentified Type~Ia remnants so that
$N_s=43-N_b$.

Table~\ref{tab:statistics} summarizes all these 
cases and gives the resulting estimates of the
binary fraction $f$ from Eqn.~\ref{eqn:binomial}. For the cases with the
interacting binaries we present the results 
assuming an underlying samples size of 23
(``no bias''), the full sample but contaminated
by 6 Type~Ia remnants (``naive Ia''), and the
full initial sample (``full bias'').  For $N_b=0$ we 
derive a 90\% confidence upper limit, and for $N_b>0$
we derive symmetric 90\% confidence limits.  This
does make the upper limits for the two cases
look qualitatively different, because the first
case has 10\% of the likelihood at higher values 
of $f$ and the second case has only 5\%.  

For the
non-interacting cases we give the results for
three mass limits, $M_* > 3$, $2$ and $1M_\odot$.
If you use the BPS models to correct for the
missing lower mass systems, the limits from
Table~\ref{tab:statistics} become
$f<0.097$, $0.098$, and $0.102$, so we will 
adopt $f \ltorder 0.10$ as our fiducial,
completeness-corrected estimate. 
For the interacting cases, we will use the
``naive Ia'' results as the fiducial values, because there must be
some Type~Ia SNR contamination and as a compromise
between ``no bias'' and ``full bias''. 
The estimates of $f=0.15$, $0.083$ and $0.082$
for the median fraction of core collapse SNRs with
interacting binaries in the ``no bias'', ``naive Ia''
and ``full bias'' are all mutually consistent given
the overall uncertainties.

Figure~\ref{fig:velos} shows the distribution of the binaries
and single NSs in transverse velocity simply using the nominal
values from Table~\ref{tab:results}.  The single NSs show the
broad distribution of velocities that drives the need for a
kick velocity produced by the ccSN explosion. The 3 HMXBs,
on the other hand, have very small transverse velocities,
and this is also true of the HMXBs from the \cite{Liu2006}
catalog.  This was noted previously
for a smaller number of systems with Hipparcos measurements
(\citealt{Chevalier1998}) and by the distances between HMXBs
and star clusters in the Magellanic Clouds (\citealt{Coe2005}).
Because these high mass stars are short lived compared to
Galactic orbital periods, the transverse velocity distributions
of the field HMXBs are likely quite similar to their natal
values.  

In Figure~\ref{fig:velos} we also show the transverse velocity
distribution of the unbound NS and the star/NS binaries from 
the BPS models.  We are measuring the transverse velocities 
relative to a rest frame defined by the local stars, while the
BPS model velocities in Figure~\ref{fig:vdist} are measured
relative to the pre-SN rest frame of the binary.  To model 
the observations, we add a velocity dispersion of $10$~km/s to 
each velocity component from the BPS models to include the 
typical random velocities of young stars relative to the local 
mean motion (see, e.g., \citealt{Binney1998}).  The parameters
of the BPS models were tuned to agree with studies of NS kicks,
so it is not surprising that the match to the velocities of
the unbound NS in SNRs is quite good.  The agreement with the
velocities of both the binaries in the SNRs and the more general
HMXB population is also good even though the parameters were not
optimized to produce the agreement.  If the extra velocity 
dispersion is not included, the model distribution has too many
low velocity stars compared to the observed distribution
(compare Figures~\ref{fig:vdist} and \ref{fig:velos}).  

\section{Discussion}
\label{sec:conclude}

In a sample of 23 SNRs with compact remnants and 26 where
none have been confidently identified, we find three
surviving stellar binaries.  All three are interacting,
with two NS wind interaction systems, and one BH Roche
accretion system. This implies
(1) that the median fraction of SNRs from ccSNe with non-interacting
stellar companions is $f < 0.10$; (2) that the median fraction with 
interacting BH companions is $f=0.038$ ($0.008<f <0.10$);
(3) that 
the median fraction with interacting NS companions is 
$f=0.060$ ($0.019< f < 0.14$); and (4) and the median fraction with
either is $f=0.83$ ($0.032 < f < 0.17$).  These are 
all 90\% confidence intervals.  These are
our fiducial estimates, but the variant cases discussed 
in \S\ref{sec:results} and Table~\ref{tab:statistics}
are all broadly consistent.

The observational estimates can be
compared to the BPS estimates in Table~\ref{tab:models}.
In the BPS models, the fraction of initial binary systems
that leave a star plus compact object binary after a
ccSNe is 6\%, with nearly equal numbers of NS and BH systems.
We cannot directly compare this with \cite{Renzo2018} because
they do not subdivide the surviving bound systems
by type.  If we consider all bound systems, they find that 
11\% of SN leave some kind of binary, while we find 15\%.
Finally, we have to dilute the expected number of SNRs
containing stellar binaries for the binary fraction.
If the initial binary fraction is $F_0 = 0.84$ based
on \cite{Moe2016}, then 5\% of SNRs from ccSNe should
contain a surviving stellar binary.  This is consistent
with the observations.  

It does seem surprising that 
all three binaries in the sample are interacting.  Our crude estimate
in \S\ref{sec:theory} is that very few surviving binaries
should be close enough to immediately begin (Roche) mass transfer
after the SN, which makes SS~433 somewhat of a statistical
anomaly.  The collision of the debris with the companion
star can lead to a rapid expansion of the outer layers of
the star (e.g., \citealt{Liu2015}), which would increase the
probability of interactions while also implying that the star 
is out of equilibrium.  The two interacting NS binaries 
are less problematic. Wind interaction systems can be more
widely separated, and it seems plausible that interactions 
with the SN debris can lead the secondary to have stronger 
winds than normal.

The counterpart to searching for surviving binaries is to search
for disrupted binaries.  These should be much more common than
surviving binaries.  In our BPS models, 23\% of the initial
binary population become unbound star plus compact object
binaries after the explosion.  If the binary fraction is 84\%,
this implies that 19\% of SNRs from ccSNe should contain a
surviving, unbound stellar companion. \cite{Renzo2018} find
a higher central fraction (58\%).  The range for this fraction
cannot be directly estimated from \cite{Renzo2018}, 
but it is likely large enough to
encompass our value.   \cite{Kochanek2018} 
considered 3 core collapse SNRs (Cas~A, the Crab and SN~1987A),
finding that there could be no unbound stellar 
companions with mass ratios $q \gtorder 1$.  This implies
a 90\% confidence limit on the fraction with unbound stellar
companions of $ f< 0.44$ that is consistent with the BPS
models.  Four candidates for unbound stellar companions
have been identified in the SNRs G074.0$-$08.5
(Cygnus Loop, \citealt{Boubert2017}), G089.0$+$04.7 
(HB~21, \citealt{Boubert2017}), G180.0$-$01.7 (S147, \citealt{Dincel2015}, 
\citealt{Boubert2017}), and G205.5$+$00.5 (Monoceros Loop, \citealt{Boubert2017}).
Since the NS in G205.5$+$00.5 is in an HMXB (\citealt{Hinton2009}), 
it seems unlikely that the identification of a disrupted binary companion 
in this SNR can be correct. Cas~A, the Crab, G180.0$-$01.7 and
G205.5$+$00.5 are included in the present study, and if we view this
as the detection of one unbound companion in a sample of 4
SNRs, then the fraction of SNRs containing unbound binaries
is $f=0.31$ ($0.076 < f < 0.66$).  In total, \cite{Boubert2017}
looked at 10 SNRs, and if all 3 candidate companions other than
the one in G205.5$+$00.5 are real, and we assume the search was
``complete'', then $f=0.32$ ($0.14 < f < 0.56$).  In short, 
while the results are all very tentative, the implications of
these studies are all broadly consistent with the expectations
from BPS models (e.g., \S2, \citealt{Renzo2018}).    

The three binaries present in the sample have absolute magnitudes 
of $M_V \sim -7$ (SS~433), $-5$ (HESS~J0632$+$057) and $-6$ 
(1FGL~J1018.67$-$5856), respectively.  These are bright enough
to be identifiable in nearby galaxies ($18$-$20$~mag at
1~Mpc and $23$-$25$~mag at 10~Mpc) modulo the problem of
dust formed in the ejecta heavily obscuring them in 
young ($<10^2$~year) SNRs (\citealt{Kochanek2017}).  Their X-ray fluxes 
would be too faint for current X-ray observatories to detect them
at these distances.

Finally, we should discuss the limitations to the present study
and how they might be reduced.  The optical and near-IR 
photometry to search for the stellar counterparts is not
a major limitation.  All of the weak upper limits
in Table~\ref{tab:results} are cases where we have relied
on survey data rather than deep targeted observations.
There would be little difficulty in obtaining new observations
to drive all the upper mass limits to be significantly less 
than a Solar mass.  Uncertainties
due to extinction are already modest, and replacing all the
optical limits with near-IR limits would eliminate
extinction uncertainties as a consideration.

The properties of the SNRs are more problematic. The 
uncertainties in distance are generally more important
than any uncertainties in extinction, particularly if all
optical limits are replaced by near-IR limits.  Since
most of the SNRs contain X-ray or $\gamma$-ray sources
rather than radio pulsars, there is only modest room
to improve matters with VLBI parallaxes.  Aside from
the actual binaries, the compact objects are too faint
(if detected at all) for Gaia parallaxes.  Optical spectroscopy
of blue stars with Gaia parallaxes could be used to
better constrain the distances to the SNRs because the
SNR produces absorption lines in the spectra of background
stars. This has been used to search for single
degenerate companions in the Type~Ia Tycho remnant
(e.g., \citealt{Ihara2007}). Blue stars are required because
the absorption features all lie shortward of 4000\AA,
which means this approach will be limited by
foreground extinction.

It would also be helpful to have determined the type of
explosion, Type~Ia or core collapse, that produced the
explosion.  Type~Ia and ccSNe remnants can be distinguished 
by their degree of symmetry (e.g., \citealt{Lopez2011}), the 
relative abundance of iron and oxygen in the SNR 
(e.g., \citealt{Vink2012}, \citealt{Katsuda2018}),
or by the local stellar population (e.g., \citealt{Badenes2009},
\citealt{Jennings2012}, \citealt{Auchettl2018}). However,
there appears to be no broad survey attempting to classify
the Galactic SNRs by SN type.  

A more challenging problem is that the detectable life time
of an SNR depends on the nature of the explosion (e.g., 
\citealt{Truelove1999}, \citealt{Patnaude2017}, \citealt{Sarbadhicary2017}). 
This will be correlated with the explosion energy and ejected mass,
and this depends in turn on the binary status of the 
progenitor, since mass transfer can modify the properties
of the exploding star.  This problem can probably only be 
explored by models combining the predicted properties of 
the explosions with models for the evolution of the SNRs.
Given the large statistical uncertainties of our present
analysis, this seems unlikely to be an important problem
at present. 

Finally, there is the problem of doing a complete survey for
compact remnants in Galactic SNRs, both for the 26 SNRs
in Appendix~\ref{sec:nocompact} 
and more generally.  The primary
problem is simply that at fainter flux levels there begin to
be multiple X-ray sources projected on the remnant with too
little flux to be well-characterized given the exposure times.
Unbound NS can be identified using X-ray proper motions on
decade time scales, since the motion is $\simeq 0\farcs2$
per decade for a typical 250~km/s NS velocity at a distance
of 3~kpc (e.g., \citealt{Holland2017}).  Longer observations 
to obtain adequate X-ray spectra would also make it relatively 
easy to classify the candidates.  {\it We did see a dangerous
propensity in the literature to reject X-ray sources with
optical emission as candidates --  as should be clear from
this paper, one should expect some NS to have optical 
counterparts that are simply a normal, non-interacting 
star in a surviving binary!  }

It does appear that surviving binaries, like 
unbound stellar companions (\citealt{Renzo2018}), generally
have low (tens of km/s) transverse velocities and should be
located close to the point of explosion.  After $t = 1000 t_3$~years,
a star moving at $v = 10 v_{10}$~km/s has moved only $0\farcs7 t_3 v_{10} d_3^{-1}$
for a distance of $d =3000 d_3$~kpc.  In many cases, the
biggest uncertainty in the location of a surviving binary
will be the difficulty in determining the center of explosion
from the SNR (see, e.g., \citealt{Holland2017}) rather than
the distance of the binary from the center.  That the velocities
are so low should also greatly simplify searches for unbound
companions (e.g., \citealt{Kochanek2018}, \citealt{Boubert2017}).

\section*{Acknowledgments}

CSK thanks J. Beacom, T. Holland-Ashford, L.~Lopez, K. Stanek, 
M. Pinsonneault, and T. Thompson for discussions and comments.   
CSK is supported by NSF grants AST-1515876 and
AST-1515927.  
K. Belczynski acknowledges support from the Polish National Science Centre
(NCN) grants: 2015/19/B/ST9/01099 and 2013/10/M/ST9/00729.

\appendix
\section{SNRs With Identified Compact Objects}
\label{sec:accepted}

\begin{itemize}

% N(H)>10^22
%\item{} G007.5$-$01.7 contains PSR~J1809$-$2332 which was first identified
%  as an X-ray source (\citealt{Braje2002}) and then as a $\gamma$-ray 
%  pulsar (\citealt{Abdo2009}).  

\item{} G034.7$-$00.4 contains the radio pulsar PSR~J1856$+$0113 (\citealt{Wolszczan1991}) 
  which is not a known binary in the ATNF Pulsar Catalog, nor is it listed as
  having a proper motion measurement (\citealt{Manchester2005}).
  The absorption of $N(H) = 0.9$ to $1.7 \times 10^{22}$~cm$^{-2}$ 
  (\citealt{Shelton2004}) only marginally meets our selection criteria.
  For an estimated distance of $3.0\pm0.3$~kpc (\citealt{Ranasinghe2018}),
  the PanSTARRS extinction of $E(B-V) \simeq 1.5$ agrees with the 
  $E(B-V) \simeq 1.5$ to $2.9$ estimated from the X-ray absorption.
  There appear to be no directed optical or near-IR searches for the pulsar,
  but it has no counterpart in either PanSTARRS or UKIDSS. 
  We adopted limits of $g>23.3$, $r>23.2$, $i>23.1$, $z>22.3$ and $y>21.3$~mag
  for PanSTARRS, and  $J>19.8$, $H>19.0$ and $K>18.1$~mag for UKIDSS.

\item{ } G039.7$-$02.0 contains the interacting compact object binary SS~433
  (for a review, see \citealt{Margon1984}).  The SNR was identified prior
  to the discovery of SS~433.  
  The X-ray flux is of order $4 \times 10^{-12}$~erg~cm$^{-2}$~s$^{-1}$.
  \cite{Brinkmann1996} find $N(H) = (5.5\pm1.5) \times 10^{21}$~cm$^2$,  
  corresponding to $E(B-V) \simeq 0.9 \pm 0.3$.
  \cite{Marshall2013} estimate that the distance is $4.5\pm0.2$~kpc,
  while \cite{Blundell2004} find $5.5\pm0.2$~kpc.  Either estimate is
  consistent with the Gaia DR2 parallax of $0.2161\pm0.0626$~mas.
  The PanSTARRS extinction estimates extend only to $\sim 3.7$~kpc,
  where they reach $E(B-V) \simeq 1.7$ and \cite{Margon1984}
  cites  $E(B-V) \simeq 2.6$.  The optical fluxes of
  $V \sim 14.2$ and $B \sim 16.3$~mag (e.g., \citealt{Margon1984})
  and the 2MASS fluxes of $J=9.4$, $H=8.7$ and $K_s=8.2$~mag   
  are all dominated by accretion luminosity.  If we treat any of 
  these fluxes as upper limits, we find that they allow
  essentially any stellar mass.  In practice, the goal has
  been to identify some spectroscopic signatures of the 
  companion, with \cite{Hillwig2008} estimating that
  the companion mass is $(12.3 \pm 3.3)M_\odot$.  \cite{Lockman2007}
  estimate that SS~433 has a 3D peculiar velocity of $\sim 35$~km/s.
  Using 1660 Gaia DR2 stars with parallaxes within $1\sigma$ of
  SS~433 to define the local rest frame,
  SS~433's
  proper motions of ($-2.85\pm 0.10$, $-4.57\pm 0.10$)~mas/year are
  statistically consistent with the mean and dispersion of 
  $(-1.80\pm 2.29$, $-4.13 \pm 2.68$)~mas/year found for these
  nearby stars.  Formally, we find a 2D transverse velocity of 
  $25 \pm 8$~km/s relative to the mean motion of these stars.  

%\item{ } G049.2$-$00.7 is associated with the 
%  X-ray source CXO~J192318.5$+$1403035, which has an unabsorbed X-ray
%  flux of $1 \times 10^{-12}$~erg~cm$^{-2}$~s$^{-1}$ (\citealt{Koo2005}). 
%  \cite{Sasaki2014} find $N(H) = (1.5-2.1)\times 10^{22}$~cm$^2$, roughly corresponding to
%  $E(B-V) \simeq 3.1 \pm 0.6$.
%  \cite{Tian2013} estimate a distance of $4.3$~kpc, while \cite{Ranasinghe2018} find $5.4$~kpc.
%  The PanSTARRS extinction estimates are only valid to $4.3$~kpc, where $E(B-V) \simeq 1.7$.
%  There are no optical counterparts in PanSTARSS, and no near-IR counterparts
%  in UKIDSS (REF).  There is a $J=19.5$, $H=18.2$, $K=17.0$ UKIDSS source 1\farcs8
%  from the X-ray position, but this distance is much larger than typical uncertainties
%  in CXO astrometry.  Based on the position of the NS relative to the center
%  of the SNR, \cite{Koo2005} estimate a transverse velocity of $\sim 360$~km/s.

\item{ } G065.7$+$01.2 contains the resolved pulsar wind nebula (PWN) DA~495 and a central object  J1952.2$+$2925 with 
  a thermal X-ray spectrum and an unabsorbed X-ray flux of order $10^{-12}$~erg~cm$^{-2}$~s$^{-1}$
  (\citealt{Arzoumanian2004}). \cite{Arzoumanian2008} find $N(H) \simeq (2-7) \times 10^{21}$~cm$^2$
  corresponding to $E(B-V) \simeq 0.7 \pm 0.5$.  The amount of absorption very roughly
  constrains the distance to be between $1$ and $5$~kpc (\citealt{Karpova2016}), which is consistent
  with the PanSTARRS dust models. There is a PanSTARRS/UKIDSS source within $1\farcs1$ 
  of the X-ray position having magnitudes $g=20.21\pm0.02$, $r=19.05\pm0.01$, $i=18.48\pm 0.02$
  $z=18.19\pm0.02$, $y=17.77\pm0.02$, $J=16.54\pm0.01$, $H=16.05\pm0.01$ and $K=15.82\pm 0.03$.
  We adopt these as the limiting magnitude for any counterpart to the X-ray source.
  We could find no estimates of the transverse velocity.

\item{ } G069.0$+$02.7 (CTB~80) is associated with the radio pulsar PSR~B1951+32 (\citealt{Kulkarni1988}).
  It is not reported to be a binary pulsar in the ATNF Pulsar Catalog (\citealt{Manchester2005}).
  It has an X-ray flux of roughly $4\times 10^{-13}$~erg~cm$^{-2}$~s$^{-1}$ and $N(H) \simeq 3.0 \times 10^{21}$~cm$^{-2}$,
  corresponding to $E(B-V)  \simeq 0.5$~mag (\citealt{Safi1995}).  
  \cite{Butler2002} identify two possible optical counterparts to the pulsar
  with F547M magnitudes of $24.26 \pm 0.30$ and $24.54 \pm 0.12$~mag. Improved
  astrometry by \cite{Moon2004} is only consistent with the brighter of these
  two candidates.  The PanSTARRS dust maps give $E(B-V)=0.73$ for a distance 
  of $2.4$~kpc, consistent with the estimate from $N(H)$.  \cite{Koo1993}
  roughly estimate the distance to CTB~80 to be $d\simeq 2$~kpc, while
  \cite{Leahy2012} estimate a distance of $1.5_{-0.4}^{+0.6}$~kpc. \cite{Kulkarni1988}
  estimate $1.4$~kpc from the dispersion measure to the pulsar.  
  \cite{Migliazzo2002} measure a proper motion of ($-11.5\pm3.1$,$-26.3\pm 3.7$)~mas/year
  corresponding to $240 \pm 40$~km/s at a distance of 2~kpc.

% N(H) too high
%\item{ } G076.9$+$01.0 is associated with the X-ray and radio pulsar PSR~J2022$+$3842 
%  (\citealt{Arzoumanian2011}) which is not a known pulsar binary in the ATNF
%  pulsar catalog (\citealt{Manchester2005}). 
%  X-ray spectral models give $N(H)=(1.7\pm0.3) \times 10^{22}$~cm$^{-2}$ (\citealt{Arzoumanian2011}), 
%  corresponding to $E(B-V)=2.9\pm 0.5$, which is somewhat higher than the PanSTARRS
%  estimate of $E(B-V)=2.1$ at their distance estimate of $10$~kpc.  There is no
%  UKIDSS (\citealt{Lucas2008}) source at the position of the pulsar, which implies an upper limit on
%  any companion of $J>19.8$, $H>19.0$ and $K>18.1$~mag, and there is no PanSTARRS (\citealt{Chambers2016})
%  source, which implies upper limits of $g>23.3$, $r>23.2$, $i>23.1$, $z>22.3$ and $y>21.3$~mag.
%  Of these limits, the H band proves to be the most restrictive.  
%  We could find no estimates of the transverse velocity.

\item{ } G078.2$+$02.1 is associated with the $\gamma$-ray pulsar PSR~J2021$+$4026, which is
  not a known pulsar binary in the ATNF pulsar catalog (\citealt{Manchester2005}).  The
  X-ray flux is of order $9 \times 10^{-13}$~erg~cm$^{-2}$~s$^{-1}$.
  \cite{Landecker1980} estimate a distance of $1.5 \pm 0.5$~kpc.
  The X-ray absorption implies $N(H)=6.4_{-1.8}^{+0.8} \times 10^{21}$~cm$^2$ (\citealt{Hui2015}), corresponding to $E(B-V) = 1.1 \pm 0.2$,
  while the PanSTARRS estimates span $E(B-V)=1.5$ to $1.6$ over the estimated distance range.
  \cite{Weisskopf2006} and \cite{Trepl2010} found no optical counterpart to the pulsar. Stronger limits
  are set by the absence of a counterpart in IPHAS ($r>21.2$, $i>20.0$, \citealt{Barentsen2014}) 
  and UKIDSS (\citealt{Lucas2008}).  H band provides the strongest limit on the mass of any
  stellar counterpart.  Based on the position of the NS relative to the center of the SNR,
  \cite{Hui2015} estimate a transverse velocity of $\sim 550$~km/s.

\item{ } G106.3$+$02.7 is associated with the radio pulsar PSR~J2229$+$6114 (\citealt{Halpern2001b}),
  which is not a known pulsar binary in the ATNF pulsar catalog (\citealt{Manchester2005}). 
   The X-ray flux is $\sim 2 \times 10^{-12}$~erg~cm$^{-2}$~s$^{-1}$.
  X-ray absorption estimates give $N(H)=(6.3\pm 1.3) \times 10^{21}$~cm$^2$
  (\citealt{Halpern2001}), corresponding to $E(B-V) = 1.1 \pm 0.2$~mag. Based
  on the X-ray absorption, \cite{Halpern2001} argue for a distance of $\sim 3$~kpc,
  and at this distance, PanSTARRS estimates a consistent extinction of 
  $E(B-V) \simeq 1.0$~mag.  \cite{Halpern2001b} obtain a flux limit of $R<23$~mag for any 
  optical counterpart.  We could find no estimates of the transverse velocity.

\item{ } G109.1$-$01.0 is associated with AXP 1E~2259$+$586 (\citealt{Gregory1980}). The X-ray flux
  is $\sim 2 \times 10^{-11}$~erg~cm$^{-2}$~s$^{-1}$.  X-ray absorption estimates give
  $N(H)=(0.93\pm0.04) \times 10^{22}$~cm$^2$ (\citealt{Patel2001}),
  corresponding to $E(B-V)=1.60\pm0.07$.  \cite{Verbiest2012} estimate a 
  distance of $4.1\pm0.7$~kpc and the PanSTARRS extinction estimates 
  for this distance are lower, at $E(B-V) \simeq 0.9$ to $1.0$.
  \cite{Hulleman2000} report optical detection limits of $R>25.7$
  and $I>24.3$ while \cite{Hulleman2001} report $R>26.4$, $I>25.6$, 
  $J>23.8$ and $K_s=21.7\pm0.2$~mag.  The IR flux is variable and
  correlated with the X-ray flux (\citealt{Tam2004}), so the IR detection 
  should be viewed as an upper limit on any stellar companion.
  \cite{Tendulkar2013} measure a proper motion of ($-6.4\pm 0.6$, $-2.3\pm 0.6$)~mas/year
  corresponding to a transverse velocity of $157\pm 17$~km/s for a distance of 
  $3.2$~kpc.

\item{ } G111.7$-$02.1 (Cas~A) is associated with the X-ray source 
  CXOU~J232327.9$+$584842 discovered in the Chandra Observatory's first
  light observations (\citealt{Tananbaum1999}). It has an unabsorbed 
  X-ray flux of flux $2 \times 10^{-12}$~erg~cm$^{-2}$~s$^{-1}$ with $N(H) \simeq (0.5-1.5) \times 10^{22}$~cm$^2$,
  corresponding to $E(B-V) \simeq 1.7 \pm 0.9$.  \cite{Reed1995} find a
  distance of roughly $3.4\pm 0.3$~kpc.  The PanSTARRS extinction estimate
  at this distance is $E(B-V) \simeq 1.2$~mag.
  \cite{Fesen2006} find no counterpart to the X-ray
  source to (STIS/50CCD) $R>28$, (F110W) $J>26.2$ and (F160W) $H>24.6$~mag.
  Note that \cite{Kochanek2018} and later \cite{Kerzendorf2018} had 
  previously concluded that Cas~A was not a binary at the time of death,
  including becoming an unbound binary. The distance of the NS from the
  expansion center of the SNR implies a transverse velocity of $\simeq 330$~km/s
  for a distance of $3.4$~kpc.

\item{ } G114.3$+$00.3 is associated with radio pulsar PSR~B2334$+$61 (\citealt{Dewey1985})
   and it is not a known pulsar binary in the ATNF pulsar catalog (\citealt{Manchester2005}).
   It has an X-ray flux of $7 \times 10^{-14}$~erg~cm$^{-2}$~s$^{-1}$, and  X-ray absorption estimates give
   $N(H) = (0.2-1.0) \times 10^{22}$~cm$^2$ (\citealt{McGowan2006}), 
   corresponding to $E(B-V) \simeq 1.0 \pm 0.7$.  Following \cite{McGowan2006} we
   adopt a distance of $3.2\pm1.7$~kpc, and over this distance range the
   PanSTARRS extinction estimates run from $E(B-V)\simeq 0.5$ to $1.0$~mag.
   There is no counterpart to the pulsar in PanSTARRS (\citealt{Chambers2016}),
   which implies upper limits of $g>23.3$, $r>23.2$, $i>23.1$, $z>22.3$ and $y>21.3$~mag.
   The $z$ band limit is the most constraining.
   \cite{Hobbs2004} measured a proper motion of ($-1 \pm 18$, $-15 \pm 16$)~mas/year.
   For 1705 stars with parallaxes consistent with this
   distance estimate, we find a mean and dispersion in the proper motions of
   ($-2.15 \pm 2.40$, $-1.10 \pm 1.60$)~mas/year. This leads to a weak 
   estimate of the transverse velocity of $212 \pm 268$~km/s. 
   \cite{Boubert2017} identify TYC~4280-562-1 in this SNR as a candidate, disrupted
   binary star.

\item{ } G119.5$+$10.2 (CTA~1) is associated with the X-ray source RX~J0007.0+7303
  (\citealt{Halpern2004}) which was later found to be a $\gamma$-ray (\citealt{Abdo2008})
  and X-ray (\citealt{Lin2010}) pulsar. The
  X-ray flux is $2 \times 10^{-13}$~erg~cm$^{-2}$~s$^{-1}$. \cite{Pineault1993} estimate a distance
  of $1.4\pm0.3$~kpc. \cite{Slane1997} find $N(H)= 2.8_{-0.5}^{+0.6} \times 10^{21}$~cm$^2$
  implying $E(B-V) \simeq 0.5 \pm 0.1$, consistent with the PanSTARRS estimate of 
  $E(B-V) \simeq 0.3$.  \cite{Mignani2013} obtained optical limits
  on any counterpart to the X-ray source of $V>26.9$ and $r>27.6$~mag.
  \cite{Slane2004a} estimated a transverse velocity of $\sim 450$~km/s for a
  distance of $1.4$~kpc based on the offset of the NS from the geometric center of the SNR.

\item{ } G130.7$+$03.1 (3C58, SN~1181) is associated with the 
  X-ray pulsar PSR~J0205$+$6449 (\citealt{Murray2002}). 
  The X-ray flux is $9 \times 10^{-13}$~erg~cm$^{-2}$~s$^{-1}$. 
  \cite{Slane2004} find $N(H)=(4.5\pm0.1)\times 10^{21}$~cm$^2$,
  corresponding to $E(B-V) \simeq 0.77 \pm 0.02$, while the PanSTARRS estimates are
  $E(B-V) \simeq 0.5$ to $0.7$.  \cite{Roberts1993} estimate a kinematic (HI) distance
  of $3.2$~kpc, while \cite{Camilo2002} obtain $4.5_{-1.2}^{+1.6}$~kpc based on the 
  dispersion measure to the pulsar.
  \cite{Moran2013} report optical detections
  of $g \simeq 27.4 \pm 0.2 $, $r \simeq 26.2\pm 0.3$ and $i \simeq 25.5\pm0.2$~mag 
  but interpret it as emission from the pulsar.
  \cite{Bietenholz2013} measure a proper motion of ($-1.40\pm0.16$, $0.54\pm 0.58$)~mas/year.
  Based on 1612 stars having parallaxes consistent
  with the compact object, we find a mean and dispersion in the proper motions
  of ($-0.71 \pm 1.79$, $0.17 \pm 1.39$)~mas/year. This is very similar to the
  proper motion of the pulsar, leading to a transverse velocity of $17\pm 9$~km/s
  that is somewhat lower than the estimate of $35 \pm 6$~km/s from \cite{Bietenholz2013}.
  The pulsar does lie close to the center of the SNR, as would be expected for such
  a low transverse velocity.

\item{ } G180.0$-$01.7 is associated with the radio pulsar PSR~J0538$+$2817
  (\citealt{Anderson1996}) and it is not a known pulsar binary in the ATNF pulsar catalog 
  (\citealt{Manchester2005}). The
  X-ray flux is $1.7 \times 10^{-12}$~erg~cm$^{-2}$~s$^{-1}$. The VLBI parallax ($0.68\pm0.15$~mas) 
  distance to the
  pulsar is $1.5_{-0.3}^{+0.4}$~kpc (\citealt{Ng2007}).  \cite{Ng2007} also find
  $N(H) \simeq (2.7\pm0.3)\times 10^{21}$~cm$^2$, implying $E(B-V) \simeq 0.5 \pm0.1$,
  a little lower than the PanSTARRS estimate of $E(B-V) \simeq 0.65$.
  There is no PanSTARRS counterpart to the pulsar.
  \cite{Ng2007} measure a proper motion of ($-23.53\pm0.16$, $52.59\pm0.13$)~mas/year.
  Based on 1573 stars with parallaxes consistent with the compact 
   object, we find a mean and dispersion in the proper motions
  of ($1.04 \pm 2.57$, $-2.86 \pm 3.29$)~mas/year.  This implies a transverse
  velocity of $425 \pm 94$~km/s, consistent the \cite{Ng2007} estimate of
  $400_{-73}^{+114}$~km/s.  This is the SNR where \cite{Dincel2015} and
  \cite{Boubert2017} identify HD~37424 as a candidate disrupted binary
  companion.

\item{ } G184.6$-$05.8 (Crab, SN~1054) is associated with the Crab radio
  pulsar PSR~B0531$+$21 (\citealt{Staelin1968}), which is not a known pulsar
  binary (\citealt{Manchester2005}).  
  \cite{Willingale2001} find $N(H)=(3.45 \pm 0.02) \times 10^{21}$~cm$^2$,
  corresponding to $E(B-V) \simeq 0.587\pm 0.003$ in reasonable agreement
  with the PanSTARRS estimate of $E(B-V) \simeq 0.4$.  We adopt a distance
  of $2.0\pm 0.5$~kpc (\citealt{Kaplan2008}). \cite{Sandberg2009} 
  give magnitudes of $V=16.66\pm0.03$, $R=16.17\pm0.02$, $I=15.65\pm 0.02$,
  $z=15.39\pm 0.05$, $J=14.83\pm 0.03$, $H=14.28\pm0.02$ and $K_s=13.80\pm 0.01$~mag.
  This is all non-thermal emission from the pulsar (see the review of the 
  Crab by \citealt{Hester2008}), and non-detection of a binary in the 
  pulsar timing represents a much stronger mass limit on any companion.
  In the local standard of rest of the Crab, \cite{Kaplan2008} estimate that the
  proper motions are ($-11.8\pm 2.0$, $4.4\pm 2.0$)~mas/year, corresponding
  to a transverse velocity of $120$~km/s at a distance of $2$~kpc.

\item{ } G189.1$+$03.0 (IC~443) is associated with the NS CXO~J061705.3$+$222127 and 
  the PWN G189.22$+$2.90 (\citealt{Keohane1997}).  The total (PWN$+$NS) X-ray flux 
  is $5 \times 10^{-12}$~erg~cm$^{-2}$~s$^{-1}$. \cite{Fesen1984} estimates a distance to the remnant of
  $1.5$ to $2.0$~kpc, which is supported by \cite{Welsh2003}.  \cite{Gaensler2006}
   find $N(H) = (7.2\pm0.6)\times 10^{21}$~cm$^2$, corresponding to 
  $E(B-V) \simeq 1.2 \pm 0.1$, while the PanSTARRS estimates for the
  assumed distances are lower at $E(B-V) \simeq 0.8$. There is no
  PanSTARRS source corresponding to the X-ray source.
  \cite{Swartz2015} estimate a transverse velocity of $400$-$600$~km/s based
  on the separation of the NS from the geometric center of the SNR
  for a distance of $1.5$~kpc.

\item{ } G205.5$+$00.5 (Monoceros Loop) is associated with the 
  $\gamma$-ray source HESS~J0632$+$057.  The SNR was identified 
  prior to the discovery of the $\gamma$-ray source.
  \cite{Hinton2009} also
  identified it as an X-ray source and suggested that it was in a 
  binary with the massive star MWC~148.
  The high energy emissions are believed to be due to 
  interactions between a pulsar and the stellar wind 
  (see the review by \citealt{Dubus2013}).
  \cite{Hinton2009} find $N(H) = (3.1\pm0.3)\times 10^{21}$~cm$^2$
  corresponding to $E(B-V) \simeq 0.53\pm 0.05$.  
  \cite{Odegard1986} estimated an HI distance of $1.6$~kpc while \cite{Zhao2018}
  find $2.0$~kpc.  The PanSTARRS extinction estimate for these 
  distances is $E(B-V) \simeq 0.6$.
  MWC~148 has a Gaia DR2 parallax of $0.3625\pm 0.0440$~mas that
  is consistent with the SNR distance estimates.
  The star has 2MASS magnitudes of $J=7.64\pm 0.02$, 
  $H=7.39 \pm 0.05$ and $K_s = 6.97\pm 0.02$, APASS
  magnitudes of $V=9.07$, $B=9.63$, $g=9.34$, $r=8.78$ and
  $i=8.70$~mag, and \cite{Neckel1980} find $V=9.17$,
  $B=9.72$, and $U=9.17$.  \cite{Bongiorno2011} found a $321\pm 5$~day
  periodicity in the X-ray emission, a period consistent with
  the radial velocity variations of the star (\citealt{Cesares2012},
  \citealt{Moritani2018}). \cite{Aragona2010} obtain a spectroscopic
  temperature of $\log T \simeq 30000$~K and combining this with
  SED fits estimate that the mass of the star is $13 M_\odot < M < 19 M_\odot$.
  When we fit these magnitudes to the PARSEC models, using the average B and V magnitudes,
  the priors in Table~\ref{tab:results}, uncertainties of $0.1$~mag, and terms
  for the Gaia parallax and the spectroscopic temperature ($T=29000\pm2000$~K)
  we find a best fit 
  model with $M = 29 M_\odot$ ($20 M_\odot < M < 36 M_\odot$, for $\Delta\chi^2=4$), 
  $T \simeq 29200$~K ($24100 < T < 32100$), $d \simeq 2.26$~kpc
  ($1.97 < d < 2.58$),
  and $E(B-V) \simeq 0.96$ ($0.92 < E(B-V) < 0.98$).  The fits are not 
  great ($\chi^2=34$), presumably because
  we have not taken into account any emission from the disk of this
  B0pe star (see \citealt{Aragona2010}), but are broadly consistent with the
  prior estimates.  Using 1260 stars having parallaxes
  consistent with MWC~148 to define a local rest frame, we find a 
  2D transverse velocity of $3 \pm 1$~km/s.  The proper motion of MWC~148,
  ($-0.08\pm0.08$, $-0.55\pm0.07$)~mas/year is statistically consistent with the
  local mean and dispersion of ($-0.05\pm 1.78$, $-0.80\pm2.18$)~mas/year.
  \cite{Boubert2017} identify HD~261393 as a candidate disrupted binary
  companion in this SNR.  However, it was identified by only one of their
  two methods, and the existence of the HMXB makes the identification
  unlikely.

\item{ } G260.4$-$03.4 (Puppis A) was first associated with
  an X-ray source (\citealt{Petre1996}) that was later found to be
  an X-ray pulsar (PSR~J0821$-$4300, \citealt{Gotthelf2009}).
  It has an X-ray flux of $3 \times 10^{-12}$~erg~cm$^{-2}$~s$^{-1}$.
  \cite{Reynoso2017} estimate a kinematic distance to the SNR of
  $1.3 \pm 0.3$~kpc. \cite{Hui2006} find $N(H)=(3.7\pm0.1)\times 10^{21}$~cm$^2$
  corresponding to $E(B-V)=0.63\pm0.02$.  This cannot be checked in
  PanSTARRS as it lies outside of the PS1 survey region. \cite{Mignani2009}
  find no optical counterpart down to $5\sigma$ limits of $B\simeq 27.2$,
  $V\simeq 26.9$ and $I \simeq 25.6$~mag.
  \cite{Becker2012} measure a proper motion of ($-64\pm12$, $-31\pm13$)~mas/year.
  Based on 1857 stars having parallaxes consistent
  with the compact object, we find a mean and dispersion in the proper motions
  of ($-3.32 \pm 2.94$, $4.00 \pm 3.58$)~mas/year.  This implies a transverse
  velocity of $433 \pm 126$~km/s. This is lower than the estimate of
  $672\pm115$~km/s by \cite{Becker2012} only because they use a larger
  distance of $2$~kpc.

\item{ } G263.9$-$03.3 (Vela) is associated with the Vela pulsar (PSR~J0835$-$4510, \citealt{Large1968}).
  The PWN has an X-ray flux of $5 \times 10^{-11}$~erg~cm$^{-2}$~s$^{-1}$, and \cite{Pavlov2001}
  find $N(H) = (3.0\pm 0.3) \times 10^{20}$~cm$^2$, implying a negligible
  extinction of $E(B-V) \simeq 0.05$.  The VLBI parallax
  distance to the pulsar is $0.29 \pm 0.02$~kpc (\citealt{Dodson2003}).
  \cite{Nasuti1997} measured optical fluxes of $U=23.38\pm0.15$,
  $B=23.89\pm0.15$, $V=23.65\pm 0.10$ and $R=23.93\pm 0.20$ which
  is believed to be emission from the pulsar. \cite{Dodson2003} measure
  a proper motion of ($-49.68\pm0.06$, $29.9\pm 0.1$)~mas/year.
  Based on 1414 stars having parallaxes consistent
  with the compact object, we find a mean and dispersion in the proper motions
  of ($-5.46 \pm 10.18$, $3.93 \pm 11.65$)~mas/year.  This implies a transverse
  velocity of $70 \pm 8$~km/s, consistent with the estimate of $62 \pm 2$~km/s
  by \cite{Dodson2003}, where their smaller uncertainty does not include a contribution
  from the kinematic model.

\item{ } G266.2$-$01.2 (Vela Jr.) is associated with the X-ray source CXOU~J085201.4$-$461753
  (\citealt{Pavlov2001b}).  It has a flux of $2 \times 10^{-12}$~erg~cm$^{-2}$~s$^{-1}$, and
  \cite{Pavlov2001b} find $N(H)= (3\pm1) \times 10^{21}$ corresponding 
   to $E(B-V) \simeq 0.5 \pm 0.2$.
   \cite{Allen2015} adopt a distance of $0.7 \pm0.2$~kpc.
    \cite{Mignani2007} identify a possible counterpart in the near-IR
   with $R >25.6$, $J>22.6$, $H\simeq 21.6 \pm 0.1$ and $K_s \simeq 21.4 \pm0.2$~mag.
   We could find no estimates of the transverse velocity.

\item{ } G284.3$-$01.8 is associated with the $\gamma$-ray source
  1FGL~J1018.6-5856, and it was identified as an HMXB by \cite{Corbet2011}.
  There is also an X-ray pulsar, PSR~J1016$-$5857, on the edge or just
  outside the remnant (\citealt{Camilo2001}), which we will ignore.
  The SNR was identified prior to the discovery of the binary.
  Like HESS~J0632$+$057, the high energy emission is believed to
  be due to interactions between a pulsar and the stellar wind
  (see the review by \citealt{Dubus2013}).
  The binary has an X-ray flux of $2 \times 10^{-12}$~erg~cm$^{-2}$~s$^{-1}$.
  \cite{Williams2015} find $N(H) = (9.0 \pm 0.9) \times 10^{21}$~cm$^2$, 
  corresponding to $E(B-V) \simeq 1.5 \pm 0.2$.  The Gaia DR2 parallax is
  $0.153\pm 0.025$~mas, consistent with the previous distance estimate 
  of $5.4_{-2.1}^{+4.6}$~kpc by \cite{Napoli2011}.  The star is
  a main sequence O6 star (\citealt{Corbet2011}, \citealt{Waisberg2015}), and \cite{Napoli2011}
  successfully model the spectral energy distribution with 
  $R=10.1R_\odot$, $T=38900$~K, $L=10^{5.3} L_\odot$ for
  $E(B-V)=1.34$ and $d=5.4$~kpc.   
  Both \cite{Waisberg2015} and \cite{Strader2015} favor a NS as the
  compact companion. The binary period is 16.6 days (\citealt{Fermi2012}).
  The 2MASS magnitudes are $J=10.44$, $H=10.14$ and $K_s=10.02$ and 
  the AAVSO magnitudes are $B=13.64$, $V=12.68$, $g=13.16$, $r=12.29$,
  and $i=11.85$.  We fit these magnitudes assuming $0.1$~mag uncertainties 
  with the distance and extinction priors from Table~\ref{tab:results} and then 
  added terms for the Gaia DR2 parallax and the spectroscopic
  temperature ($T = 38000 \pm 2000$~K for 06, \citealt{Martins2005})
  to the goodness of fit.  This leads to a mass estimate of
  $40 M_\odot$ ($29M_\odot < M < 61M_\odot$) with distances and extinctions of
  $5.0 < d < 9.2$~kpc and $1.33 < E(B-V) < 1.35$, respectively,  
  consistent with \cite{Napoli2011}. If we include the Swift
  photometry from \cite{Fermi2012}, we are unable to obtain
  a good fit to the SED ($\chi^2=356$ instead of $2.8$), but 
  the best models favor lower 
  masses, temperatures and extinctions ($M \simeq 20M_\odot$,
  $T \simeq 10^4$~K, $E(B-V) \simeq 1.0$).
  Using 1562 stars with consistent parallaxes,
  we estimate a transverse velocity of $39 \pm 7$~km/s.
  The proper motion of the binary, ($-6.41\pm0.05$,$2.21\pm0.05$)~mas/year,
  is statistically consistent with the motions of nearby stars, which have
  a mean and dispersion of ($-5.51\pm 1.91$, $3.10\pm 1.67$)~mas/year.

\item{ } G291.0$-$00.1 is associated with the PSR candidate CXOU~J111148.6$-$603926
  and a PWN with an X-ray flux of $4 \times 10^{-13}$~erg~cm$^{-2}$~s$^{-1}$ (\citealt{Slane2012}). 
  \cite{Slane2012} find
  $N(H) = (6.7 \pm 0.7) \times 10^{21}$~cm$^2$, corresponding to
  $E(B-V) \simeq 1.1 \pm 0.1$.  The distance to the SNR
  is not well established, with a minimum distance of $3.5$~kpc and a typical
  scaling to 5~kpc (see the discussion in \citealt{Slane2012}).
  There is no optical counterpart in Gaia DR2, which we interpret
  at $G<22$, $G_{BP}<20$ and $G_{RP}<20$~mag. 
  \cite{Holland2017} used proper motions to derive a transverse
  velocity of $303 \pm 130$~km/s for a distance of $5$~kpc.

\item{ } G296.5$+$10.0 is associated with the X-ray PSR~J1210$-$5226 (\citealt{Zavlin2000}).
    It has an X-ray flux of $2\times 10^{-12}$~erg~cm$^{-2}$~s$^{-1}$, and
   \cite{deLuca2004} find $N(H)= (1.3\pm 0.1)\times 10^{21}$~cm$^2$, corresponding
   to $E(B-V) \simeq 0.22 \pm 0.02$.
   \cite{Giacani2000} estimate a distance of $2.1_{-0.8}^{+1.8}$~kpc.
    \cite{deLuca2004} also found no optical counterparts down to $R>27.1$
    and $V>27.3$~mag. 
   We could find no estimates of the transverse velocity.

\item{ } G320.4$-$01.2 is associated with the X-ray pulsar PSR~J1513$-$5908
  (\citealt{Seward1982}). The X-ray flux is $6 \times 10^{-12}$~erg~cm$^{-2}$~s$^{-1}$, and
  \cite{Yatsu2005} find $N(H) = (0.86 \pm 0.09) \times 10^{22}$~cm$^2$, corresponding
  to $E(B-V) \simeq 1.5 \pm 0.2$ which cannot be checked with PanSTARRS.
  \cite{Gaensler1999} estimate a distance of $5.2\pm1.4$~kpc.
  \cite{Kaplan2006} identify a possible counterpart with $R \simeq 25.6 \pm 0.3$, 
  $J>20.7$, $H\simeq 20.6\pm 0.2 $ and $K_s \simeq 19.4 \pm 0.1$~mag, but argue that 
  the emission is likely from the pulsar.  
   We could find no estimates of the transverse velocity.

\item{} G332.4$-$00.4 contains the central compact object 1E~161348$-$5055 (\citealt{Tuohy1980}).
  The X-ray flux shows a $\simeq 6.7$~hour periodicity.  
  \cite{Reynoso2004} estimate a distance  
  of $3.1$~kpc, and \cite{Frank2015} find column densities of $N(H)= (0.6-1.4) \times 10^{22}$~cm$^{-2}$
  corresponding to $E(B-V) \simeq 1.0$ to $2.4$.  
  The field is crowded, and \cite{deLuca2008} find no compelling near-IR counterpart to the
  X-ray source.  If we use their near-IR limits of $H>23$ and $K_s >22.1$~mag, then
  the mass limit is $<0.1M_\odot$.  If we identify the counterpart as their star \#5
  with $H=21.43 \pm0.01$ and $K=19.22\pm 0.21$, then the limit is $<0.2M_\odot$. 
  However, \cite{deLuca2008} note that the colors of star \#5 are those of a more
  distant, more heavily extincted source.  We treat this as a non-detection.

\end{itemize}

\section{SNRs Without Identified Compact Objects}
\label{sec:nocompact}

\begin{itemize}

\item{} G007.7$-$03.7 has no reported association, although there are
   X-ray observations of the remnant (\citealt{Zhou2018}).

\item{} G025.1$-$02.3 has no reported association.  It is not clear
   whether it has ever been searched for either an X-ray source or 
   a radio pulsar.

% A_V too high no N(H)
%\item{ } G030.7$+$0.10 is not associated with any radio, X-ray or $\gamma$-ray
%  counterpart. 

\item{} G032.8$-$00.1 has an unrelated pulsar, PSR~J1853$-$0004, and an unrelated
  background variable X-ray source (2XMM~J185114.3$-$000004, \citealt{Bamba2016}).

\item{} G038.7$-$01.3 contains four faint X-ray sources (\citealt{Huang2014})
  but they are all below our flux limit of  $10^{-13}$~erg~cm$^{-2}$~s$^{-1}$.  

% A_V too high no N(H), no distance
%\item{ } G043.9$+$01.6 (and G042$+$0.6) have been been associated with 
%  SGR~1900$+$14 ($F_X \sim 2 \times 10^{-12}$~erg~cm$^{-2}$~s$^{-1}$). However, proper motion measurements
%  show that it is not associated with either SNR (\citealt{deLuca2009},
%  \citealt{Tendulkar2012}).

\item{} G053.6$-$02.2 has no associated sources, but has been observed in X-rays (e.g., \citealt{Broersen2015}).

\item{} G055.7$+$3.4 meets our selection criteria, but is little studied.  The pulsar PSR~J1921$+$2153
   lies on the edge of the shell and is probably unrelated (\citealt{Bhatnagar2011}).

% distance too high
%\item{ } G067.7$+$01.8 has no X-ray sources brighter than
%  $5 \times 10^{-14}$~erg~cm$^{-2}$~s$^{-1}$ (\citealt{Hui2009}), which is below our threshold. We note
%  that \cite{Hui2009} reject 2 of the four faint X-ray sources they consider
%  because they have optical counterparts, which should not be done because
%  the optical emission could well be from a surviving binary companion.

\item{} G065.3$+$05.7 meets our selection criteria.  The pulsar PSR~J1931$+$30 is nearby, 
   but probably is not not associated. No candidate NS has been identified in X-ray observations
   (\citealt{Kaplan2006}).

\item{ } G074.0$-$08.5 (Cygnus Loop) contains a $2 \times 10^{-13}$~erg~cm$^{-2}$~s$^{-1}$
   X-ray source, but its properties make it unlikely to be associated with the 
   SNR unless it has a very high ($\sim 2000$~km/s) transverse velocity (\citealt{Katsuda2012}).  

% distance too high
%\item{} G074.9$+$01.2 is associated with the PSR candidate CXOU J201609.2+371110 and
%   the PWN  G74.94$+$1.11 but the observed X-ray flux is only $4 \times 10^{-14}$~erg~cm$^{-2}$~s$^{-1}$
%   (\citealt{Matheson2013}).
  
\item{} G082.2$+$05.3 has no associated sources, but has been observed in X-rays (e.g., \citealt{Mavromatakis2004}). 

\item{} G085.4$+$00.7 has a number of superposed X-ray sources (\citealt{Jackson2008}), some of 
   which may slightly exceed our $10^{-13}$~erg~cm$^{-2}$~s$^{-1}$ flux limit (the paper does
   not report fluxes). None are identified as a likely NS.

\item{} G085.9$-$00.6 has a number of superposed X-ray sources (\citealt{Jackson2008}), some of 
   which may slightly exceed our $10^{-13}$~erg~cm$^{-2}$~s$^{-1}$ flux limit (the paper does
   not report fluxes). None are identified as a likely NS.
`
\item{} G089.0$+$04.7 has X-ray observations (e.g., \citealt{Pannuti2010}), but no reported
   association.

\item{ } G093.3$+$06.9 contains a candidate PWN, but the flux is 
    only $4\times 10^{-14}$~erg~cm$^{-2}$~s$^{-1}$ (\citealt{Jiang2007}).

\item{ } G116.9$+$00.2 (CTB~1) has been associated with the X-ray source
  RX~J0002$+$6246, with an X-ray flux of $3 \times 10^{-13}$~erg~cm$^{-2}$~s$^{-1}$.  However,
  \cite{Esposito2008} conclude that it is actually X-ray emission from
  a foreground star. The Gaia DR2 parallax of the star, $3.137\pm0.038$~mas, confirms
  this conclusion.

\item{ } G127.1$+$00.5 has no X-ray sources brighter than 
  $5 \times 10^{-14}$~erg~cm$^{-2}$~s$^{-1}$ and this source has
  the  X-ray/optical flux ratio of a star (\citealt{Kaplan2004}).  

\item{ } G132.7$+$01.3 (HB3) is near the radio pulsar PSR~J0215$+$6218.
   However, \cite{Lorimer1998} do not believe the two are associated
   because the spin down age is orders of magnitude larger than the
   estimated age of the SNR.

\item{} G156.2$+$05.7 has no associated X-ray source (\citealt{Kaplan2006}).

\item{ } G160.9$+$02.6 (HB9) is unlikely to be associated with 
  SGR~0501$+$4516, as it lies outside the rim of the SNR
  (e.g., \citealt{Gaensler2008}). The SNR is also probably 
  unrelated to PSR~B0459$+$47/J0502$+$4654/B0458$+$46 (see, e.g., \citealt{Kaplan2006}).

\item{} G166.0$+$04.3 has X-ray observations (e.g., \citealt{Bocchino2009}), but no 
  associated source.

% no modern X-ray
%\item{ } G213.0$-$00.6 might be associated with the ROSAT source 1RXS~J065049.7$-$003220 
%  (\citealt{Stupar2012}), but there is almost no other information.  The nearby, 
%  bright ($V \sim 6$~mag), foreground binary star HIP~32851 makes this a poor 
%  location to search for counterparts in any case. 

% no distance
%\item{ } G279.0$+$01.1 has no associated X-ray source or radio pulsar.

% no distance
%\item{ } G289.7$-$00.3 has no associated X-ray source or radio pulsar.

\item{} G296.1$-$00.5 has X-ray observations (e.g., \citealt{Castro2011}), but no
  associated source.

\item{} G309.2$-$00.6 contains an X-ray source, but it is an unrelated, foreground Be star
  (\citealt{Safi2007}).

% no distance
%\item{G315.4$-$02.3} contains several X-ray sources but their fluxes are 
%  too low at $5 \times 10^{-14}$~erg~cm$^{-2}$~s$^{-1}$ (\citealt{Mignani2012}). 

\item{} G326.3$-$01.8 is associated with an X-ray point source/PWN with a flux  
  of $5 \times 10^{-14}$~erg~cm$^{-2}$~s$^{-1}$ (\citealt{Yatsu2013}),
  which is below our flux threshold.

\item{} G330.0$+$15.0 (Lupus loop) has no X-ray sources above 
  $\sim 1 \times 10^{-13}$~erg~cm$^{-2}$~s$^{-1}$ (\citealt{Kaplan2006}).

\item{} G332.5$-$05.6 has no associated source.

\item{ } G343.1$-$02.3 is associated with the radio pulsar
   PSR~J1709$-$4429 (\citealt{Johnston1992}), but it lies near
  the very edge of the remnant.   It is difficult to reconcile
  its properties with the transverse velocity required to 
  reach the edge of the SNR (see the discussions in \citealt{Dodson2002}
  and \citealt{Romani2005}).
  The X-ray flux is $2 \times 10^{-13}$~erg~cm$^{-2}$~s$^{-1}$.

\item{} G343.0$-$0.60 has no associated source.

\end{itemize}

\section{Rejected SNRs}
\label{sec:rejected}

\begin{itemize} 

\item{} G004.5$+$06.8 (Kepler) is a Type~Ia SN (e.g., \citealt{Reynolds2007}). 

\item{} G011.1$+$00.1 is probably a remnant in the background of the closer
    PSR~J1809$-$1917 (see \citealt{Kargaltsev2007}).  The Manitoba SNR
    catalog uses the distance to the pulsar, but our distance selection
    criterion is meant to apply to the SNR.  Hence, neither the compact
    object nor the SNR should be included in our statistics. 

\item{} G120.1$+$01.4 (Tycho) is a Type~Ia SN (e.g., \citealt{Krause2008}).

\item{} G315.4$-$02.3 is generally identified as a Type~Ia remnant (e.g., \citealt{Williams2011},
   \citealt{Yamaguchi2014}).  In particular, the Fe K$\alpha$ line emission of the remnant
   is typical of Type~Ia SNRs and incompatible with that of core collapse SNRs (\citealt{Yamaguchi2014}).
   The distance is $2.5 \pm 0.5$~kpc (\citealt{Helder2013}) and \cite{Castro2013} find
   $N(H) = 6 \times 10^{21}$~cm$^{-2}$ corresponding to $E(B-V) \simeq 1.0$.  
   \cite{Gvaramadze2017} propose that the X-ray source identified as $[\hbox{GV2003}]$N 
   in \cite{Gvaramadze2003} is a surviving low mass NS binary associated with a core
   collapse SNe that produces the SNR despite lying at the remnant edge.  The source has
   $V=20.69\pm 0.02$, $i=18.77 \pm 0.08$, $z=18.29 \pm 0.06$,  $J=16.71 \pm 0.17$,
   $H = 15.87 \pm 0.14$, $K_s=15.73 \pm 0.18$~Vega mag and a spectroscopic
   temperature of $T_*=5100 \pm 200$~K.  With our standard fitting procedures,
   we find reasonably good fits for a $0.9 M_\odot < M_* < 1.2 M_\odot$ main
   sequence star, consistent with the estimate by \cite{Gvaramadze2017}.  
   The abundances of this G star appears to be anomalous, which
   \cite{Gvaramadze2017} interpret as contamination by ejecta from a calcium-rich
   SN.  It is also difficult to reconcile the typical distance of Ca-rich SN from 
   their host galaxies (e.g., \citealt{Kasliwal2012}) with a massive star origin for
   the Ca-rich transients.
   Unfortunately, the Gaia DR2 parallax of the source ($\pi = 8.0 \pm 1.0$~mas,
   or a distance of $125\pm15$~kpc) seems to be clearly wrong since since it
   also has a fit statistic of $\chi^2=2392$.  In our view, this binary seems
   most likely to be a chance projection.

\item{} G327.6$+$14.6 is the remnant of SN~1006, which was probably a Type~Ia SN
   (see the review by \citealt{Vink2012}).

\end{itemize}

\end{document}